\documentclass[12pt]{article}
\usepackage[inline,shortlabels]{enumitem}
\usepackage{float}
\usepackage{mathtools}
\usepackage{graphicx}
\usepackage[round]{natbib}
\usepackage[margin=1in]{geometry}
\usepackage{tikz}
\usepackage[english]{babel}
\usepackage{longtable}
\usepackage{color}
\usepackage{amssymb,amsmath,amsthm}
\usepackage{multirow}
\usepackage[titletoc,title]{appendix}
\usepackage{authblk}
\usepackage{setspace}
\usepackage{dsfont}
\usepackage[OT1]{fontenc}
\usepackage{refcount}
\usepackage{booktabs}
\usepackage{arydshln}
\graphicspath{ {./figs/} }

\usepackage{nameref,zref-xr} % nameref before zref-xr
\zxrsetup{toltxlabel} % allows to use the \ref command as usual

\usepackage{subfig}
\usepackage[inline]{trackchanges}
\addeditor{Ivan}
\addeditor{Nima}
\addeditor{Kara}

\newtheorem{theorem}{Theorem}
\AtEndDocument{\refstepcounter{theorem}\label{finalthm}}
\AtEndDocument{\refstepcounter{equation}\label{finaleq}}

{
  \theoremstyle{definition}
  
}
{
  \theoremstyle{definition}
  
}
{
  \theoremstyle{definition}
  
}

\AtEndDocument{\refstepcounter{lemma}\label{finallemma}}

\DeclareMathOperator{\expit}{expit}

\DeclareMathOperator{\logit}{logit}

\renewcommand{\P}{\mathsf{P}}

\newcommand{\p}{\mathsf{p}}
\newcommand{\q}{\mathsf{q}}
\newcommand{\g}{\mathsf{g}}
\newcommand{\ps}{\mathsf{t}}
\newcommand{\e}{\mathsf{e}}

\newcommand{\uu}{\mathsf{u}}
\newcommand{\h}{\mathsf{h}}
\newcommand{\vv}{\mathsf{v}}
\newcommand{\rr}{\mathsf{r}}
\newcommand{\y}{\mathsf{b}}
\newcommand{\s}{\mathsf{c}}

\newcommand{\indep}{\mbox{$\perp\!\!\!\perp$}} 

\newcommand{\dd}{\mathrm{d}}

\newcommand{\one}{\mathds{1}}
 
\newcommand{\E}{\mathsf{E}}

\DeclarePairedDelimiterX{\norm}[1]{\lVert}{\rVert}{#1}

\pgfdeclarelayer{background}
\pgfsetlayers{background,main}
\usetikzlibrary{arrows,positioning}
\tikzset{
>=stealth',
punkt/.style={
rectangle,
rounded corners,
draw=black, very thick,
text width=6.5em,
minimum height=2em,
text centered},
pil/.style={
->,
thick,
shorten <=2pt,
shorten >=2pt,}
}
\newcommand{\Vertex}[2]% pos, name
{\node[minimum width=0.6cm,inner sep=0.05cm] (#2) at (#1) {$\footnotesize#2$};
}
\newcommand{\Vertexr}[2]% pos, name
{\node[rectangle, draw, minimum width=0.6cm,inner sep=0.05cm] (#2) at (#1) {$\footnotesize#2$};
}
\newcommand{\ArrowR}[3]%
{ \begin{pgfonlayer}{background}
\draw[->,#3] (#1) to[bend right=30] (#2);
\end{pgfonlayer}
}
\newcommand{\ArrowL}[3]%
{ \begin{pgfonlayer}{background}
\draw[->,#3] (#1) to[bend left=45] (#2);
\end{pgfonlayer}
}
\newcommand{\EdgeL}[3]%
{ \begin{pgfonlayer}{background}
\draw[dashed,#3] (#1) to[bend right=-45] (#2);
\end{pgfonlayer}
}

\newcommand{\Arrow}[3]%
{ \begin{pgfonlayer}{background}
\draw[->,#3] (#1) -- +(#2);
\end{pgfonlayer}
}
\newcommand{\titlepaper}{Nonparametric estimators of interventional (transported) direct and indirect effects that accommodate multiple mediators and multiple intermediate confounders.}%Efficiently transporting causal direct and indirect effects to new populations under intermediate confounding and with multiple mediators}

\date{}

\author[1]{Kara E. Rudolph}
\author[1]{Nicholas T. Williams}
\author[2]{Iv\'an D\'iaz}
\affil[1]{\small Department of
  Epidemiology, Mailman School of Public Health, Columbia University.}
\affil[2]{\small Division of Biostatistics, Department of Population
  Health, New York University Grossman School of Medicine.}
\usepackage[colorlinks,citecolor=blue,urlcolor=blue]{hyperref}
\title{\titlepaper}

\date{}

\begin{document}
\maketitle

\begin{abstract}
Mediation analysis is appealing for its ability to improve understanding of the mechanistic drivers of causal effects, but real-world data complexities challenge its successful implementation, including: 1) the existence of post-exposure
variables that also affect mediators and outcomes (thus, confounding the mediator-outcome relationship), that may also be 2) multivariate, and
3) the existence of multivariate mediators. Interventional direct and indirect effects (IDE/IIE) accommodate post-exposure variables that confound the mediator-outcome relationship, but currently,
no estimator for IDE/IIE exists that allows for both multivariate mediators and multivariate post-exposure intermediate confounders. This, again, represents
a significant limitation for real-world analyses. %, because many real-world data scenarios involve
%multivariate mediators and intermediate confounders.
We address this gap by extending two recently developed nonparametric estimators---one that estimates the IDE/IIE and another that estimates the IDE/IIE transported to a new, target population---
to allow for multivariate mediators and multivariate intermediate confounders simultaneously. We use simulation to examine finite sample performance, and apply these estimators to longitudinal data from the Moving to Opportunity trial. In the application, we walk through a strategy for
separating indirect effects into mediator- or mediator-group-specific indirect effects, while appropriately accounting for other, possibly co-occurring intermediate variables.
\end{abstract}

\section{Introduction}
Causal mediation analysis is used to delineate and estimate the causal paths by which an exposure is linked to an outcome. Consequently, mediation can be useful to understand \textit{how} an exposure exerts an effect on an outcome. For example, mediation has been used to quantify the extent to which the effect of placental abruption (a pregnancy complication) on perinatal mortality operates through preterm birth \citep{ananth2011placental}, and to quantify the extent to which use of a Section 8/ Housing Choice voucher negatively impacts the mental health of adolescent boys through features of their school environments \citep{rudolph2021helped}. (We consider the latter as our motivating example here.) In addition, estimating mediation mechanisms across subgroups with heterogeneous treatment effects can shed light on the extent to which differences in the mechanisms by which the exposure affects the outcome could contribute to the heterogeneous treatment effects. For example, this strategy identified that part of the reason why homeless and housed individuals with opioid use disorder respond differently to medications (in terms of medication effectiveness in preventing relapse) is due to differences in mediation mechanisms through adherence, illicit opioid use, depressive symptoms, and pain \citep{rudolph2021explaining}. 

Although mediation analysis is appealing for its ability to improve understanding of the mechanistic drivers of causal effects, several real-world data complexities challenge its successful implementation. The most common of these data complexities include: 1) the existence of post-exposure variables that also affect mediators and outcomes (thus, confounding the mediator-outcome relationship), henceforth referred to as \textit{intermediate confounders}, that may also be 2) multivariate, and 3) the existence of multivariate mediators. We next explain why these common data characteristics pose such a problem.

Natural direct and indirect effects (NDE/NIE) %are arguably the causal mediation estimand that 
most closely represent many mediational research questions, because they summarize individual-level causal path-specific effects \citep{robins1992identifiability,Pearl01}. For example, the NIE describes how an individual's mediator value would change under contrasting exposure levels and how those counterfactual mediator values would affect an outcome of interest. We can formalize this definition using notation, showing how the NIE and NDE decompose the average treatment effect (ATE): \begin{equation*}
\E(Y_{1, M_1} - Y_{0, M_0}) =
      \underbrace{\E(Y_{1, M_1} - Y_{1, M_0})}_{\text{natural indirect
          effect (through $M$)}} +
  \underbrace{\E(Y_{1, M_0} - Y_{0,M_0})}_{\text{natural direct effect (not through $M$)}}
\label{eq:decomp},
\end{equation*} where $A$ denotes treatment, $Y$ denotes outcome, $M$ denotes the mediator(s), $M_{A=a'}$ denotes the counterfactual mediator value(s) had treatment been set to $A=a'$, possibly contrary to fact, and where $Y_{A=a, M=M_{a'}} $ denotes the nested counterfactual outcome had treatment been set to $A=a$ and had the mediator value(s) been what they would have been under treatment $A=a'$. 

\begin{figure}[H]
  \caption{Directed acyclic graph of the structural causal model considered.}
\centering
\includegraphics[width=.5\textwidth,keepaspectratio]{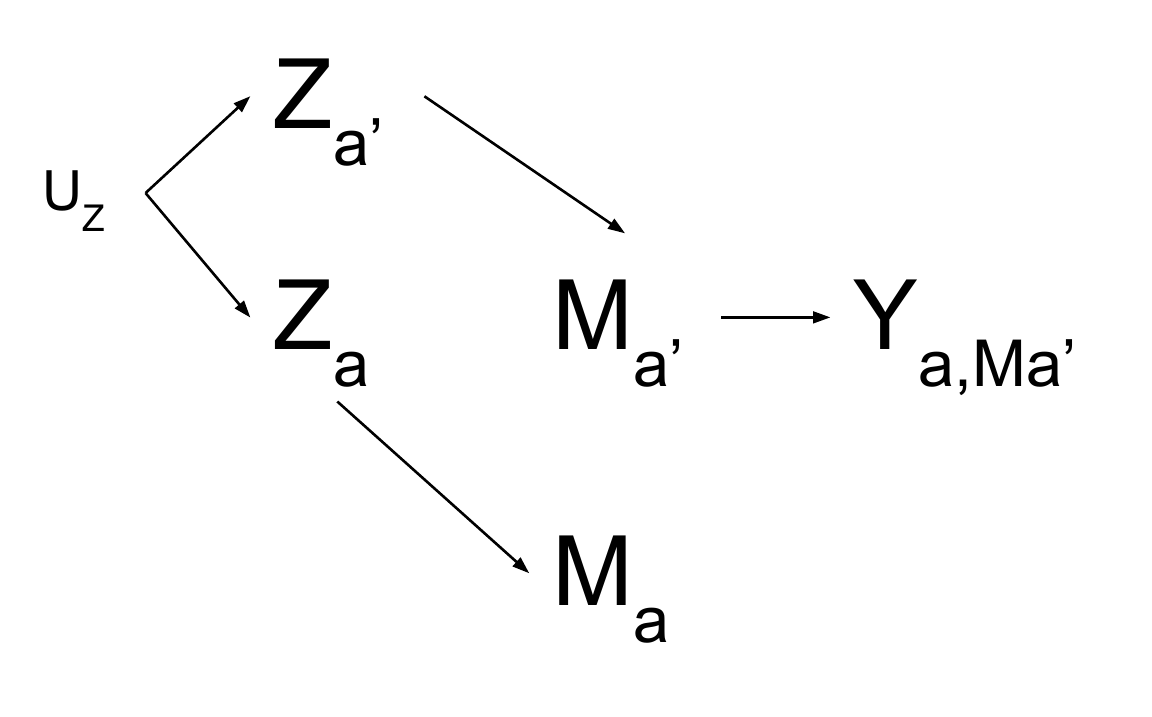}
\label{dagfig}
\end{figure}

However, the NIE and NDE are not generally point identified in the presence of a post-exposure intermediate confounder \citep{avin2005identifiability}. To give intuition for this, consider the counterfactual outcome $Y_{A=a, M_{A=a'}}$. It is defined invoking two counterfactual worlds simultaneously: 1) $A=a$, which induces the counterfactual intermediate confounder $Z_{A=a}$; and 2) $A=a'$, which induces the counterfactual intermediate confounder $Z_{A=a'}$. $Z_{A=a}$ and $Z_{A=a'}$ share unmeasured
  common causes, $U_Z$, which creates a spurious association between $M_{a'}$, and $Y_{A=a, m}$, meaning that they cannot be conditionally independent, depicted in Figure \ref{dagfig}. 
  
This is a significant limitation for real-world analyses, because post-exposure intermediate confounders are near-ubiquitous. For example, they are always present in trials where randomized assignment to a treatment influences the treatment taken. They are also present in all observational studies where the exposure influences a host of variables that precede and also influence the outcome. A subset of these variables may be the mediators of interest. Another subset that are not of interest as mediators could plausibly affect the mediators (in addition to affecting the outcomes), so would be post-exposure intermediate confounders.

Randomized interventional direct and indirect effects (IDE, IIE) have been proposed as a ``second best'' causal mediation estimand that is point identified under less restrictive assumptions that allow for the presence of post-treatment confounders of the mediator-outcome relationship \citep{vanderweele2014effect}. Instead of defining a counterfactual outcome value under a static intervention on $A=a$ and a nested counterfactual mediator value setting $A=a'$, $Y_{a,M_{a'}}$, as for the NDE/NIE, interventional direct and indirect effects define a counterfactual outcome value under a static intervention on $A=a$ but a stochastic draw from the counterfactual distribution of $M_{A=a'}$, conditional on $W$. The stochastic draw from the distribution of $M_{A=a'} \mid W$ is denoted $G_{a'}$. Then, the corresponding counterfactural outcome value is denoted $Y_{a,G_{a'}}$. The stochastic draw from the counterfactual distribution $\P(M_{a'} \mid W)$ means the assumption that $M_{a'}$ is conditionally independent of $Y_{a, m} \mid W$ is no longer required for identification, though the remaining NDE/NIE identification assumptions remain the same. Thus, the IDE/IIE can be identified in the presence of the intermediate confounders.

But, as stated above, these estimands are considered ``second best'', because they no longer summarize individual-level mediational paths. Instead, they summarize population-level mediational paths. This is a problem, because mediation is typically considered an individual-level phenomenon. %individual-level paths are what we think of when we think about causal mediation. 
Based on this premise, \citet{miles2022causal} defined the sharp mediational null as $H_0: Y^i_{a',M^i_a} = Y^i_{a',M^i_{a^\star}} \text{ for both } a'=a \text{ and } a'=a^\star$ for each $i$ in the population of interest. This sharp mediational null means that individual-level NIE is zero for everyone in the population---i.e., no one in the population experiences mediation. \citet{miles2022causal} argues that an estimand of the indirect effect should respect the sharp mediational null, with a true value of zero when no one in the population experiences mediation. While the NIE satisfies the sharp null, the IIE is not guaranteed to do so. For example, if one subset of the population experiences an effect between $A$ and $M$, and another, distinct subset experiences an effect between $M$ and $Y$ but no one has both a relationship between $A$ and $M$ and $M$ and $Y$, then the IIE could be non-null even though no individual experiences mediation. This is a theoretical problem; it is not clear to what extent it is a problem in real-world analyses. %Nonetheless, the IIE remains the only point-identified causal indirect effect estimand in the presence of a post-treatment intermediate variable.

Methods for estimating the IDE/IIE consist of parametric and nonparametric approaches \citep{vanderweele2017mediation,zheng2017longitudinal,diaz2021nonparametric,rudolph2018robust,hejazinonparametric}. Consider an observed data scenario with intermediate confounder(s) and mediators, $O=(W, A, Z, M, Y)$, defined above. Parametric estimation approaches would rely on the ability to correctly specify models for both the mediator and outcome (which would be required for a parametric marginal structural model (MSM) estimator for a single mediator) or models for the outcome, mediator, and intermediate confounder (which would be required for a parametric g-computation estimator, also for a single mediator and single intermediate confounder). Correctly specifying a single parametric model may be unlikely, and correct specification of multiple models may be even more tenuous. Multiple nonparametric estimators exist that place no restrictions on the joint distribution of $W$ and no restriction on $Y$ \citep{zheng2017longitudinal,diaz2021nonparametric,rudolph2018robust,hejazinonparametric}. Some nonparametric estimators of the IDE/IIE also allow for continuous $A$ (instead of the more typical binary $A$) \citep{hejazinonparametric}, or continuous or multivariate $Z$ or $M$, but not both \citep{diaz2021nonparametric}. Currently, no estimator for IDE/IIE exists that allows for both multivariate $Z$ and $M$. This, again, represents a significant limitation for real-world analyses, because many real-world data scenarios involve multivariate $Z$ and $M$. 

We address this gap here. Specifically, we extend a nonparametric estimator of the IDE/IIE \citep{diaz2021nonparametric} to allow for multivariate $Z$ and multivariate $M$ simultaneously. We allow for multivariate $M$ by a reparameterization similar to that proposed by \citet{diaz2021nonparametric}. However, our proposed reparameterization is modified such that we do not rely on estimating the density of $Z=z$ conditional on treatment and covariates, $\p(z \mid a', W)$. Instead, we estimate the density ratio: $\frac{\p(z \mid a', W)}{\p(z \mid a', m, W)}$. While nonparametrically estimating the density of a multivariate parameter is hard, nonparametrically estimating such a density ratio is achievable and has been done previously \citep{diaz2021lmtp}, which we describe further below. We also propose a similar extension to a related estimator for the IDE/IIE transported to a new target population, called the \textit{transported IDE and IDE} \citep{rudolph2022efficiently}, to allow for multivariate $Z$ and multivariate $M$ simultaneously. Our proposed one-step and partial targeted minimum loss-based estimators are multiply robust, efficient, and use data-adaptive, machine learning algorithms in model fitting. 

This paper is organized as follows. In Section 2, we introduce notation, the structural causal models (SCMs) we consider, definitions, and review previously established identification results. In Section 3, we give the nonparametric efficient influence function for the non-transported and transported IDE/IIE and describe our proposed one-step and TML estimators. We describe and provide results for a simulation study of each estimator's finite sample performance in Section 4. In Section 5, we apply each estimator to a motivating data example from a multi-site trial, the Moving to Opportunity Study (MTO), where families living in low-income public housing were randomized to receive a Section 8 housing voucher or not and followed for 10-15 years \citep{sanbonmatsu2011moving}. We apply the non-transported estimators to estimate the effect of randomized housing voucher receipt for young children on behavioral problems later in adolescence, possibly operating through: i) characteristics of the school environment, ii) number of school and residential moves, and iii) neighborhood poverty. We consider the mediators together as a bundle as well separately by category i), ii), or iii). When considered separately, we walk through how one can appropriately account for other, possibly co-occuring intermediate variables. We apply the transported estimators to the same MTO study and question, transporting the IIE from the Los Angeles study site to the Boston study site. Each of Sections 2-5 %is broken up into two subsections: the 
first addresses the nontransported IDE/IIE, and second, addresses the extension to the transported IDE/IIE. %; readers interested in just one type of estimator can skip the other subsection without loss of information. 
Section 6 concludes.  
%%%%%%%%%%%%%%%%%%%%%%%%%%%%%%%%%%%%%%%%%%%%%%%%%%%%%%%%%%%%%%%%%%%%%%%%%%%%%%%
\section{Notation and background on the definition and identification of (in)direct effects}

Let $O = (W, A, Z, M, Y)$ represent the observed data in the nontransported setting, and let $O = (S, W, A, Z, M, SY)$ represent the observed data in the transported setting, where $S$
denotes a binary variable indicating membership in the source
population ($S=1$) or target population ($S=0$), $W$ denotes a vector
of observed pre-treatment covariates; $A$ denotes a categorical
treatment variable; $Z$ denotes intermediate confounders (mediator-outcome confounders affected by treatment) that may be binary, continuous, or multivariate; $M$ denotes
mediators that may be binary, continuous, or multivariate; and $Y$ denotes a continuous or binary
outcome. Let $O_1, \ldots, O_n$ denote a sample of $n$
i.i.d.~observations of $O$. Note that in the transported setting, we assume the outcome is only observed for
the source population/sites, $S=1$, but we are interested in
estimating effects for the target population/site, $S=0$. 

We formalize
the definition of our counterfactual variables using the following
non-parametric structural equation model \citep[NPSEM, ][]{Pearl2009}
though other approaches may be taken. In the nontransported setting, assume the
data-generating process satisfies:
\begin{multline}\label{eq:npsem}
  W = f_W(U_W);\ A = f_A(W, U_A);\ Z=f_Z(W, A, U_Z);\\
  M = f_M(W, A, Z, U_M);\ Y = f_Y(W, A, Z, M, U_Y).
\end{multline} In the transported setting, assume the related
data-generating process:
\begin{multline}\label{eq:npsem}
  S=f_S(U_S);\ W = f_W(S,U_W);\ A = f_A(S,W, U_A);\ Z=f_Z(S,W, A, U_Z);\\
  M = f_M(S,W, A, Z, U_M);\ Y = f_Y(W, A, Z, M, U_Y).
\end{multline}
$U=(U_S,U_W,U_A,U_Z,U_M,U_Y)$ is a vector of exogenous factors, and
functions $f$ are deterministic and unknown. $\P$
denotes the distribution of $O$, and is  %, and $\Ps$ to denote the
%distribution of $(O,U)$. 
an element of the nonparametric statistical model
defined as all continuous densities on $O$ with respect to some
dominating measure $\nu$. Let $\p$ denote the corresponding
probability density function. We denote random variables with capital
letters and realizations of those variables with lowercase
letters. %We let $\E$ and
%$\E_c$ denote corresponding expectation operators, and define
 For a given function $f(o)$, $\P f = \int f(o)\dd \P(o)$.% and use $\Pn$ to denote the empirical distribution of $O_1, \ldots, O_n$.

 We use the following additional definitions.  The function $\s(a,z,m,w)$ denotes
 $\P(S=1\mid A=a,Z=z,M=m,W=w)$; $\g(a \mid w)$ denotes $\P(A=a\mid W=w)$ in the nontransport setting and
 $\P(A=a\mid W=w,S=0)$ in the transport setting; $\e(a \mid m, w)$ denotes
 $\P(A=a\mid M=M,W=w)$ or $\P(A=a\mid M=M,W=w,S=0)$; $\q(z \mid a,w)$ denotes the density of
 $Z$ conditional on $(A,W)=(a,w)$ or conditional on $(A,W,S)=(a,w,0)$; $\rr(z \mid a,m,w)$ denotes the
 density of $Z$ conditional on $(A,M,W)=(a,m,w)$ or conditional on $(A,M,W,S)=(a,m,w,0)$; $\y(a,z,m,w)$
 denotes $\E(Y \mid A = a,Z = z,M = m, W = w)$ or $\E(Y \mid A = a,Z = z,M = m, W = w, S=1)$; and $\ps$ denotes
 $\P(S=0).$
 % \begin{itemize}
% \item[] $\s(a,z,m,w)$ denotes $\P(S=1\mid A=a,Z=z,M=m,W=w)$,
%     \item[] $\g(a \mid w)$ denotes $\P(A=a\mid W=w,S=0)$, %the probability mass function
% %of $A=a$ conditional on $(W,S) = (w, 0)$, 
%     \item[] $\e(a \mid m, w)$ denotes $\P(A=a\mid M=M,W=w,S=0)$,%the probability mass function of $A=a$ conditional on
% %$(M, W, S)=(m,w,0)$, 
%     \item[] $\q(z \mid a,w)$ denotes $\P(Z=z\mid A=a,W=w,S=0)$,%the density of $Z$ conditional on $(A,W,S)=(a,w,0)$,
%     \item[] $\rr(z \mid a,m,w)$ denotes $\P(Z=z\mid A=a,M=m,W=w,S=0)$,%the density of
% %$Z$ conditional on $(A,M,W,S)=(a,m,w,0)$,
%    \item[] $\y(a,z,m,w)$ denotes $\E(Y \mid A = a,Z = z,M = m, W = w, S=1)$, and
%   \item[] $\ps$ denotes $\P(S=0).$
% \end{itemize}
 For a random variable $X$, we let $X_a$ denote the counterfactual
 outcome observed in a hypothetical world in which $\P(A=a)=1$. For
 example, we have $Z_a = f_Z(S,W,a,U_Z)$, $M_a=f_M(S,W,a,Z_a,U_M)$,
 and $Y_a=f_Y(W,a,Z_a,M_a,U_Y)$. Likewise, we let
 $Y_{a,m} =f_Y(W,a,Z_a,m,U_Y)$ denote the value of the outcome in a
 hypothetical world where $\P(A=a,M=m)=1$.

\subsection{Interventional direct and indirect effects}
\label{sec:estimand}
As described in the Introduction, the IDE and IIE have been defined and identified previously \citep{vanderweele2014effect}. We review the necessary background here. Let $G_a$ denote a random draw from the conditional distribution of
$M_a$ conditional on $(W)$. The IIE is defined as:
$\E(Y_{a', G_{a'}} - Y_{a', G_{a^{\star}}})$, and is interpreted as %As we assume $Y$ is unobserved among those with $S=0$
the population-level path from $A$ to $M$, including through $Z$, to $Y$. It is the average difference in counterfactual outcome values had all the units been treated
while varying the stochastic draw of mediator values from their counterfactual distribution
under treatment versus their counterfactual (possibly joint) distribution under no treatment, conditional on
covariates. 

The IDE is
similarly defined:
$\E(Y_{a', G_{a^{\star}}} - Y_{a^{\star}, G_{a^{\star}}})$,
and is interpreted as the population-level path from $A$ to $Y$, including through $Z$, but not through $M$. It is the average difference in
counterfactual outcome values had all the units been treated versus not, while stochastically drawing the mediators from their counterfactual distribution under
no treatment, conditional on covariates. 

%We focus on identification and estimation
%of $\theta = \E(Y_{a', G_{a^{\star}}})$. %Contrasts of $\theta$
%under the values of $a'$ and $a^*$ given %in the above definitions
%correspond to the IDE and IIE.  

Under the assumptions

\begin{enumerate}[label=(\roman*)]
\item $Y_{a,m}\indep A\mid W$,\label{ass:ncay}
\item $M_{a}\indep A\mid W$,\label{ass:ncam}
\item $Y_{a,m}\indep M\mid (A,W,Z)$,\label{ass:ncmy}
   and 
 \item positivity: $\p(w)>0$ implies $\p(a \mid w)>0$ for $a \in\{a', a^\star\}$; and $\p(w)>0, \p(z \mid a',w)>0$, and $\p(m \mid a^\star, w)>0$ imply $\p(m \mid z,a',w)>0.\label{ass:pos} $ 
 %there is a non-zero probability of assigning any level $A$ for all observed $w$; a non-zero probability of assigning any level $A$ for all $w,m$; a non-zero probability of assigning any level of $Z$  for all combinations of $W, A, M$; and a non-zero probability that $S=0$ for all combinations of $W, A, Z, M$ (referred to as the positivity assumption),\label{ass:pos}
\end{enumerate}
\citet{vanderweele2014effect} showed that each component of the IDE/IIE contrast, $\theta = \E(Y_{a', G_{a^{\star}}})$, is identified and is equal to
\begin{equation}
\theta = \int \y(a',z,m,w)\q(z\mid
  a',w)\p(m\mid a^{\star},w)\p(w)\dd \nu(w,z,m).\label{eq:thetadef}
\end{equation} Assumptions \ref{ass:ncay} - \ref{ass:ncmy} are sequential exchangeability assumptions, meaning that  conditional on $W$, there is no unmeasured confounding of
the relation between $A$ and $Y$ or $A$ and $M$; and
conditional on $(A,W,Z)$ there is no unmeasured confounding of the
relation between $M$ and $Y$. 

\subsection{Transported interventional direct and indirect effects}
\label{sec:transpestimand}
Transported interventional direct and indirect effects (transported IDE, IIE) have also been defined and identified previously \citep{rudolph2019transporting,rudolph2022efficiently}. We review the necessary background here. For transported versions of these effects, we let $G_a$ denote a random draw from the conditional distribution of
$M_a$ conditional on $(S=0,W)$. The transported IIE among those for whom
$S=0$ is defined as:
$\E(Y_{1, G_{1}} - Y_{1, G_{0}}\mid
S=0)$. The transported IDE among those for whom $S=0$ can be
similarly defined:
$\E(Y_{1, G_{0}} - Y_{0, G_{0}}\mid S=0)$. These effects are interpreted analogously as described in the above subsection, the difference being that they are effects transported from the source to the target population.

Under the assumptions
\begin{enumerate}[label=(\roman*)]
\item $Y_{a,m}\indep A\mid W$,\label{ass:ncayt}
\item $M_{a}\indep A\mid W, S=0$,\label{ass:ncamt}
\item $Y_{a,m}\indep M\mid (A,W,Z)$,\label{ass:ncmyt}
\item
  $\E(Y\mid A = a,Z = z,M = m, W = w, S=1)=\E(Y\mid A = a,Z = z,M = m,
  W = w, S=0)$,\label{ass:excht} 
   and 
 \item positivity: $\p(w \mid S=0)>0$ implies $\p(a \mid w, S)>0$ for $a \in\{a', a^\star\}$; and $\p(w \mid S=0)>0, \p(z \mid a',w, S=0)>0$, and $\p(m \mid a^\star, w, S=0)>0$ imply $\p(z \mid a',w, S=1)>0$ and $\p(m \mid z,a',w \mid S=1)>0 $  ,\label{ass:post}
\end{enumerate}
\citet{rudolph2022efficiently} showed that each component of the TIDE/TIIE contrast $\theta^T = \E(Y_{a',G_{a^\star}} \mid S=0)$ is identified and is equal to
\begin{equation}
\theta^T = \int \y(a',z,m,w, S=1)\q(z\mid
  a',w,S=0)\p(m\mid a^{\star},w, S=0)\p(w\mid S=0)\dd \nu(w,z,m).\label{eq:thetadef}
\end{equation}  
Assumption \ref{ass:excht} is commonly referred to as the ``transport assumption'', and allows us to transport or borrow
information on the outcome model from other sites \citep{rudolph2022efficiently,pearl2011transportability}. If $Y$ is observed
among those for whom $S=0$% , as is the case in our illustrative
% example
, then this assumption can be tested nonparametrically \citep{luedtke2019omnibus}.
%
%In the next section we turn our attention to a discussion of
%efficiency theory for the estimation of the statistical distribution
%$\P$.

\section{Proposed nonparametric estimators for $\theta$ and $\theta^T$}
We propose one-step and partial targeted minimum loss-based estimators (TMLEs) of $\theta$ and $\theta^T$ that are based on the efficient influence function (EIF). The so-called `one step' estimator has its name, because when the EIF estimating equation is linear, it solves it in one step. Our primary motivation for proposing estimators based on the EIF is that this allows one to use data-adaptive machine learning algorithms in model fitting while retaining the ability to compute theoretically correct standard errors and confidence intervals. In addition, these estimators are locally efficient (meaning that their variance is the lower bound of the asymptotic variance of any regular estimator of the parameter, $\theta$, under the model considered) and multiply robust (meaning that certain components of the data distribution can be inconsistently estimated but the estimator remains consistent).

\subsection{Efficient influence function for $\theta$ and $\theta^T$}
 We 
first reparameterize the previous EIFs provided in \citet{diaz2021nonparametric} and \citet{rudolph2022efficiently} for the nontransported and transported IDE/IIE, respectively.  

%The EIF characterizes the
%asymptotic behavior of all regular and efficient estimators
%\citep{Bickel97, van2002part}. In addition %to being locally efficient,
%estimators constructed using the EIF have advantages of multiple
%robustness, which means that some components of the data distribution
%(i.e., nuisance parameters) can be inconsistently estimated while the
%estimator remains consistent. The multiple robustness property also
%allows the use data-adaptive machine learning algorithms in estimating
%nuisance parameters while retaining the ability to compute correct
%standard errors and confidence intervals. This is due to fact that the
%asymptotic analysis of the estimators yield second-order bias terms in
%differences of the nuisance parameters, and therefore allow slow
%convergence rates (e.g., $n^{-1/4}$) for estimating these nuisance
%parameters.
%There are  EIF is often useful in constructing
%locally efficient estimators. Some of the most common approaches for
%this are (i) using the EIF as an estimating equation
%\citep[e.g.,][]{vanderLaan2003}, (ii) using the EIF in a one-step bias
%correction \citep[e.g.,][]{pfanzagl1982contributions}, and (iii) using
%the EIF to construct targeted minimum loss-based estimators
%\citep{vdl2006targeted, vanderLaanRose11, vanderLaanRose18}. % For
% $\theta$, the first two estimators turn out to be equivalent.

\begin{theorem}[Efficient influence function for the nontransported $\theta$]\label{theo:eif}
  For fixed $a'$, $a^{\star}$ define
%\begin{equation}
%  \begin{split}
%    \h_M(z, m, w, S=0) & =  \frac{\p(m\mid a^{\star}, w, S=0)}{\p(m\mid a', z,
%      w, S=0)}\\
%    \uu(z,w) &=\int_{\mathcal
%      M}\y(a',z,m,w, S=1)\p(m\mid a^{\star},w, S=0)\dd\nu(m)\\
%    \bar\uu(w) &=\int_{\mathcal Z}\uu(z,w)\q(z\mid a', w)\dd\nu(z)\\
%    \vv(m,w) &=\int_{\mathcal
%      Z}\y(a',z,m,w)\q(z\mid a',w)\dd\nu(z)\\
%    \bar\vv(w)&=\int_{\mathcal
%      M}\vv(m,w)\p(m\mid a^\star,w)\dd\nu(m).
%  \end{split}\label{eq:defhuv}
%\end{equation}

%\begin{lemma}[Alternative representation of the EIF]\label{lemma:aeif}
 % Define $h_Z$ as
\begin{equation}
  \begin{split}  
  \h_Z(z,m,w) &= \frac{\q(z\mid a',w)}{\rr(z\mid a',m,w)}\\
  \h_M(z, m, w)&=\h_Z(z,m,w) \frac{\g(a'\mid w)}{\g(a^{\star}\mid w)}
                    \frac{\e(a^{\star}\mid m, w)}{\e(a'\mid m,w)}\label{def:hM}\\
    \uu(z,w) &= \E\left\{\y(a',Z,M,W)\h_M(Z,M,W)\,\bigg|\,
                 Z=z,A=a',W=w\right\},\notag\\
    \bar\uu(w) &= \E\left\{\uu(Z,W)\,\bigg|\,
                 A=a',W=w\right\},\notag\\
    \vv(m,w) &= \E\left\{\y(a',Z,M,W)\h_Z(Z,M,W)\,\bigg|\, M=m,
                 A=a',W=w\right\}\notag\\
    \bar\vv(w) &= \E\left\{\vv(M,W)\,\bigg|\, A=a^\star,W=w\right\} \notag.
\end{split}%\label{eq:defhuv}
\end{equation}

The efficient influence function for $\theta$ in the
nonparametric model $M$ is equal to
\begin{align}
\label{eq:eif}
\begin{split}
D_{\P,\theta}(o) = & D_{\P,Y}(o) + D_{\P,Z}(o) + D_{\P,M}(o) + D_{\P,W}(o), \text{ where } \\
  D_{\P,Y}(o) = &\frac{\one\{a=a'\}}{\g(a'\mid w)}\h_M(z,m,w)\{y - \y(a',z,m,w)\}%\label{eq:DY}
  \\
  D_{\P,Z}(o) =  & \frac{\one\{a=a'\}}{\g(a'\mid
    w)}\left\{\uu(z,w)-\bar\uu(w)\right\}%\label{eq:Du}
            \\
  D_{\P,M}(o) =  &  \frac{\one\{a=a^{\star}\}}{ \g(a^{\star}\mid w)}\left\{\vv(m,w)-\bar\vv(
            w)\right\}%\label{eq:Dv}
            \\
            D_{\P,W}(o) = & \left\{\bar\vv(w) - \theta\right\}.
  \end{split}
\end{align}
\end{theorem}

\begin{theorem}[Efficient influence function for the transported $\theta^T$]\label{theo:eif:transp}
  For fixed $a'$, $a^{\star}$ define
%\begin{equation}
%  \begin{split}
%    \h_M(z, m, w, S=0) & =  \frac{\p(m\mid a^{\star}, w, S=0)}{\p(m\mid a', z,
%      w, S=0)}\\
%    \uu(z,w) &=\int_{\mathcal
%      M}\y(a',z,m,w, S=1)\p(m\mid a^{\star},w, S=0)\dd\nu(m)\\
%    \bar\uu(w) &=\int_{\mathcal Z}\uu(z,w)\q(z\mid a', w)\dd\nu(z)\\
%    \vv(m,w) &=\int_{\mathcal
%      Z}\y(a',z,m,w)\q(z\mid a',w)\dd\nu(z)\\
%    \bar\vv(w)&=\int_{\mathcal
%      M}\vv(m,w)\p(m\mid a^\star,w)\dd\nu(m).
%  \end{split}\label{eq:defhuv}
%\end{equation}

%\begin{lemma}[Alternative representation of the EIF]\label{lemma:aeif}
 % Define $h_Z$ as
\begin{equation}
  \begin{split}  
  \h_Z(z,m,w, S=0) &= \frac{\q(z\mid a',w,S=0)}{\rr(z\mid a',m,w,S=0)}\\
  \h_M(z, m, w, S=0)&=\h_Z(z,m,w, S=0) \frac{\g(a'\mid w)}{\g(a^{\star}\mid w)}
                    \frac{\e(a^{\star}\mid m, w, S=0)}{\e(a'\mid m,w, S=0)}\\%\label{def:hM}\\
    \uu(z,w) &= \E\left\{\y(a',Z,M,W, S=1)\h_M(Z,M,W,S=0)\,\bigg|\,
                 Z=z,A=a',W=w,S=0\right\},\notag\\
    \bar\uu(w) &= \E\left\{\uu(Z,W)\,\bigg|\,
                 A=a',W=w,S=0\right\},\notag\\
    \vv(m,w) &= \E\left\{\y(a',Z,M,W,S=1)\h_Z(Z,M,W,S=0)\,\bigg|\, M=m,
                 A=a',W=w,S=0\right\}\notag\\
    \bar\vv(w) &= \E\left\{\vv(M,W)\,\bigg|\, A=a^\star,W=w,S=0\right\}\notag.
\end{split}
\end{equation}

The efficient influence function for $\theta^T$ in the
nonparametric model $M$ is equal to
\begin{align}
\label{eq:eift}
\begin{split}
D_{\P,\theta^T}(o) = & D_{\P,Y^T}(o) + D_{\P,Z^T}(o) + D_{\P,M^T}(o) + D_{\P,W^T}(o), \text{ where } \\
  D_{\P,Y^T}(o) = &\frac{\one\{s=1,a=a'\}}{\ps\times \g(a'\mid w)}\frac{1-\s(a,z,m,w)}{\s(a,z,m,w)}\h_M(z,m,w, S=0)\{y - \y(a',z,m,w, S=1)\}%\label{eq:DY}
  \\
  D_{\P,Z^T}(o) =  & \frac{\one\{s=0,a=a'\}}{\ps\times \g(a'\mid
    w)}\left\{\uu(z,w)-\bar\uu(w)\right\}%\label{eq:Du}
            \\
  D_{\P,M^T}(o) =  &  \frac{\one\{s=0,a=a^{\star}\}}{\ps\times \g(a^{\star}\mid w)}\left\{\vv(m,w)-\bar\vv(
            w)\right\}%\label{eq:Dv}
            \\
            D_{\P,W^T}(o) =  & \frac{\one\{s=0\}}{\ps}\left\{\bar\vv(w) - \theta^T\right\}.
  \end{split}
\end{align}
\end{theorem}

\subsection{Estimation of nuisance parameters}
The EIFs, given above, are functions of what are called ``nuisance parameters''. $D_{\P,\theta}$ is a function of nuisance parameters $\eta=\{\g, \e, \q, \rr, \h_Z, \h_M, \y, \uu, \bar\uu, \vv, \bar\vv\}$, and $D_{\P,\theta^T}$ is a function of nuisance parameters $\eta^T=\{\g, \e, \q, \rr, \h_Z, \h_M, \y, \s, \uu, \bar\uu, \vv, \bar\vv\}$. 

All of the above nuisance parameters may be estimated by regression,
except the density ratios $\h_Z$ and $\h_M$. For example, $\g$ can be estimated by a regression of $A$ on $W$. Parameters $\{\e, \q, \rr, \y, \s \}$ can be estimated similarly. To estimate $\{\uu, \bar\uu, \vv, \bar\vv\},$ we treat the quantity to the left of the conditioning statement as a pseudo-outcome, and regress it on the variables to the right of the conditioning statement, generating predictive values. For example to estimate $\uu$, we multiply predicted values of $\y(a',z,m,w)$ and $\h_M(z,m,w)$ together and regress the resulting estimates on $W, A, Z$. We then generate predicted values from that regression model, setting $A=a'$. One could use a parametric regression model, like logistic regression, or a nonparametric, data-adaptive model using machine learning algorithms in model fitting. We use a data-adaptive approach in what follows and in the software we propose. Specifically, we use SuperLearner, which is an ensemble of machine learning algorithms used to fit each regression, where the algorithms are weighted (via a convex combination) to minimize the 10-fold cross-validated prediction error \citep{van2007super}.

Lastly, we estimate the density ratios $\{\h_Z, \h_M\}$. It suffices to
estimate $\h_Z$ as $\h_M$ may be obtained
by solving for it in expression (\ref{def:hM}). To estimate $\h_Z$, let $\tilde M_i,\ldots,\tilde M_n$ denote $n$
independent random draws from a given distribution, for example the
empirical distribution $\P_n(m)$. Consider a dataset of size $2n$
created by duplicating all observations. This dataset is now indexed by
$\Lambda \in\{0,1\}, i \in\{1,\ldots, n\}$, where $\Lambda$ indexes the duplicated observations. In this augmented
dataset, assign $Z_i=\tilde Z_i$ to each observation with index $i$ and $\Lambda =
1$. Let $\P^\lambda$ define the distribution of the data in this
augmented dataset. Then we have the following result.
For $\theta$:
{%\footnotesize
  \begin{align*}
\h_Z(z,m,w) = &\frac{\P^\lambda(\Lambda = 1\mid A=a', Z=z, M=m,
    W=w)}{\P^\lambda(\Lambda = 0\mid A=a', Z=z, M=m,
    W=w)} \times \\ 
    &\frac{\P^\lambda(\Lambda = 0\mid A=a',M=m,
    W=w)}{\P^\lambda(\Lambda = 1\mid A=a',M=m, W=w)},
  \end{align*}}
  
and for $\theta^T$:
{%\footnotesize
  \begin{align*}
\h_{Z^T}(z,m,w) = &\frac{\P^\lambda(\Lambda = 1\mid A=a', Z=z, M=m,
    W=w, S=0)}{\P^\lambda(\Lambda = 0\mid A=a', Z=z, M=m,
    W=w, S=0)} \times \\ 
    &\frac{\P^\lambda(\Lambda = 0\mid A=a',M=m,
    W=w, S=0)}{\P^\lambda(\Lambda = 1\mid A=a',M=m, W=w, S=0)},
  \end{align*}}

where the probabilities on the right hand side of the above equation may be
estimated using any regression of $\Lambda$ on $(M,Z,W,S)$, $(M,W,S)$, $(M,Z,W)$, or $(M,W)$
in the augmented dataset. 

We propose a cross-fitted version of the one-step and partial TML estimators. Cross-fitting is a data-splitting technique that obviates the Donsker class condition that would otherwise be required for the estimator to be asymptotically normal \citep{klaassen1987consistent,zheng2011cross, chernozhukov2016double}. We perform crossfitting for estimation of all nuisance parameters, %the components of 
$\eta$ 
as follows. Let ${\cal V}_1, \ldots, {\cal V}_J$
denote a random partition of data in the index set $\{i:1,\ldots,n\}$ into $J$
prediction sets of approximately the same size such that 
$\bigcup_{j=1}^J {\cal V}_j = \{1, \ldots, n\}$. We note that we do not split the data in the augmented dataset,
$\{\Lambda:0,1;i:1,\ldots,n\}$, to ensure that the independencies
required to prove asymptotic normality of the estimators and oracle
results for cross-validation procedures remain. When we duplicate the dataset to create the augmented dataset, we include duplication of the folds. For each $j$,
the training sample is given by
${\cal T}_j = \{1, \ldots, n\} \setminus {\cal V}_j$. %First, we fit the nuisance parameters in $\eta$,
$\hat \eta_{j}$ denotes the estimator of $\eta$, obtained by training
the corresponding prediction algorithm using only data in the sample
${\cal T}_j$, and $j(i)$ denotes the index of the
validation set that contains observation $i$. We then use these fits, $\hat\eta_{j(i)}(O_i)$ in computing each efficient influence function, i.e., we compute $D_{\P, \theta}(O_i, \hat\eta_{j(i)})$. 

\subsection{One-step estimator}\label{sec:os}
The one-step estimator estimates of $\theta$ and $\theta^T$ are given by solving $\frac{1}{n} \sum_{i=1}^n D_{P,\theta}(O_i, \hat\eta_{j(i)})=0$ and\\ $\frac{1}{n} \sum_{i=1}^n D_{P,\theta^T}(O_i, \hat\eta^T_{j(i)})=0$, respectively, where the components of $\hat{\eta}$ are estimated as described above. The estimates of the nontransported and transported IDE and IIE are then given by the respective contrasts of $\theta(a', a^\star)$ and $\theta^T(a', a^\star)$. For example, the nontransported one-step IDE estimate is given by: $\hat{\theta}(1,0) - \hat{\theta}(0,0) = \frac{1}{n} \sum^n_{i=1} D_{\theta(1,0)}(O_i,\hat{\eta}_{(j)i}) - D_{\theta(0,0)}(O_i,\hat{\eta}_{(j)i}).$ The variance of these estimates is estimated as the sample variance of EIF estimates. For example, the variance of the nontransported IDE estimate is the sample variance of $D_{\theta(1,0)}(O,\hat{\eta}) - D_{\theta(0,0)}(O,\hat{\eta})$.

\subsection{Partial targeted minimum loss-based estimator}\label{sec:tmle}
We also propose a partial TMLE in an effort to improve finite sample performance, particularly in the presence of practical positivity violations \citep{petersen2012diagnosing}, which occurs when the conditional probabilities specified in Assumptions \ref{ass:pos} and \ref{ass:post} for identification of $\theta$ and $\theta^T$, respectively, are nonzero, but very small (i.e., ``practically zero''). We anticipate such finite sample issues will challenge performance of estimators for $\theta^T$ in particular; the partial TMLE may have advantages over the one-step estimator in these cases. Thus, below, we describe how to implement this estimator in terms of $\theta^T.$ Implementation for $\theta$ would proceed similarly.

Specifically, we propose a TMLE for the $Y$ component of the EIF, $D_{\P,Y^T}$, because this component is expected to be the most variable due to the weights. Thus, this partial TMLE targets $\y(a^{\prime},z,m,w,S=1)$.

First, we bound $Y$ in $[0,1]$, as described in \citet{gruber2010targeted}.
Let $\hat{\y}(a^{\prime},z,m,w,S=1)$ be an initial estimate of
$\y(a^{\prime},z,m,wS=1)$. 
We then update this initial estimate of $\hat{\y}$ using
covariate
\[\hat C_\y(a^{\prime},Z,M,W) =
  \frac{1- \hat \s(a,Z,M,W)}{\hat \s(a,Z,M,W)}\frac{\hat \h_M(Z,M,W,S=0)}{\hat \ps \times \hat \g(a^{\prime} \mid W)}\] in a logistic regression of $Y$ with
$\logit \hat{\y}(a^{\prime},Z,M,W,S=1)$ as an offset, among the subset
for which $S=1 and A=a^{\prime}$. Let $\hat{\epsilon}_\y$ denote the MLE
fitted coefficient on $\hat C_\y(a^{\prime},Z,M,W)$. The targeted (i.e.,
updated) estimate %of $\hat{m}$
is given by
\[\logit\tilde{\y}(a^{\prime},z,m,w) = \logit
  \hat{\y}(a^{\prime},z,m,w,S=1) + \hat{\epsilon}_\y \hat C_\y(a^{\prime},z,m,w).\]
As an alternative algorithm, we could use 
$\hat C_\y(a^{\prime},Z,M,W)$ as weights in an intercept-only
weighted logistic regression model. $\tilde{\y}(a^{\prime},z,m,w) $ would then be the resulting predicted value of the fitted model setting $A=a^{\prime}$.
 
We replace $\hat{\y}$ with $\tilde{\y}$ and
iterate the above until the score equation
$n^{-1}\sum_i\{D_{\tilde{\eta}, Y^T}(O_i) \}=0$ is solved up to a factor of
$(\sqrt{n}\log(n))^{-1}$. This iterating process and stopping
criterion ensures that the efficient influence function is solved up
to $n^{-1/2}$ 
and mitigates risk of overfitting. 

The other components of the EIF in Equation \ref{eq:eift} are solved as for the one-step estimator, described in Section \ref{sec:os}.
The variance of the partial TMLE estimates is estimated as the sample variance of EIF estimates. 

An R package to implement both estimators is included: %\url{blinded for review}.
\url{https://github.com/nt-williams/HDmediation/tree/main/HDmediation}.

\section{Simulation}

We conducted a simulation study to both verify correct implementation of the above R package as well as to examine the estimator's performance in finite samples. We considered two data-generating mechanisms for each of the nontransported and transported scenarios, specified in Section A1 of the online appendix.  In the first data-generating mechanism for each $\theta$ and $\theta^T$, all variables are binary---this simple scenario allows us to best verify correct implementation. The second data-generating mechanism includes multiple $Z$ and multiple $M$ to reflect the scenario where this estimator would be of practical use.

We conducted 500 simulations for sample sizes $n\in \{500,1000,10000\}$ and under correct specification of the nuisance parameters in $\eta, \eta^T$. For the scenario in which all variables are binary, we fit nuisance parameters using a saturated generalized linear model--excluding $\g$, which was fit with a mean model. For the second data generating mechanism, we fit each of the nuisance parameters using an ensemble \citep{van2007super} of a main-effects GLM, a GLM with all two-way interactions, a saturated GLM, and a saturated GLM with $\ell1$ penalization; again, fitting $\g$ with a mean model. Simulations for the second data-generating mechanism were also cross-fit with 5-folds. We considered estimator performance in terms of absolute bias, absolute bias scaled by $\sqrt{n}$, and 95\% confidence interval (CI) coverage. 

Table \ref{tab:simtheta} shows simulation results for the nontransported IDE and IIE, and Table \ref{tab:simthetat} shows simulation results for the transported IDE and IIE comparing the one-step and partial TML estimators under correct specification of all nuisance parameters.

We see in Table \ref{tab:simtheta} that, as expected, the estimators are unbiased. There is a small amount of bias in the data-generating mechanism with multivariate $M$ and $Z$ that decreases with increasing sample size. We also see relatively good performance in terms of 95\% CI coverage, with coverage of approximately 95\% in the binary $M$ and $Z$ data-generating mechanism, and slightly lower coverage of the indirect effect in the multivariate $M$ and $Z$ data-generating mechanism. Coverage appears to be slightly better for the partial TMLE as compared to the one-step estimator in this scenario. 

\begin{table}[H]
\centering
\caption{Simulation results comparing one-step and partial TML estimators for the nontransported interventional indirect and direct effects under two data-generating mechanisms: one with a single, binary $M$ and a single, binary $Z$; and another with multiple, binary $M$s and multiple, binary $Z$s.}
\label{tab:simtheta}
\footnotesize
\begin{tabular}{rcccccc}
\toprule
\multicolumn{1}{c}{ } & \multicolumn{3}{c}{Direct effect} & \multicolumn{3}{c}{Indirect effect} \\
\cmidrule(l{3pt}r{3pt}){2-4} \cmidrule(l{3pt}r{3pt}){5-7}
$n$ & $|\text{Bias}|$ & $\sqrt{n} \times |\text{Bias}|$ & 95\% Cov. & $|\text{Bias}|$ & $\sqrt{n} \times |\text{Bias}|$ & 95\% Cov.\\
\midrule
\multicolumn{7}{l}{\textbf{Binary $M$ and $Z$}} \\[2pt]
\hdashline\noalign{\vskip 1ex}
\multicolumn{7}{l}{One-step} \\[2pt]
500 & 0.00 & 0.02 & 0.94 & 0.00 & 0.03 & 0.95\\
1000 & 0.00 & 0.03 & 0.94 & 0.00 & 0.00 & 0.93\\
5000 & 0.00 & 0.06 & 0.94 & 0.00 & 0.02 & 0.94\\
10000 & 0.00 & 0.01 & 0.95 & 0.00 & 0.02 & 0.94\\[2pt] 
\hdashline\noalign{\vskip 1ex}
\multicolumn{7}{l}{Partial TMLE} \\[2pt]
500 & 0.00 & 0.09 & 0.95 & 0.00 & 0.02 & 0.95\\
1000 & 0.00 & 0.00 & 0.94 & 0.00 & 0.02 & 0.95\\
5000 & 0.00 & 0.02 & 0.93 & 0.00 & 0.04 & 0.95\\
10000 & 0.00 & 0.07 & 0.93 & 0.00 & 0.03 & 0.94\\
\midrule
\multicolumn{7}{l}{\textbf{Multivariate $M$ and $Z$}} \\[2pt]
\hdashline\noalign{\vskip 1ex}
\multicolumn{7}{l}{One-step} \\[2pt]
500 & 0.01 & 0.18 & 0.93 & 0.00 & 0.03 & 0.91\\
1000 & 0.01 & 0.27 & 0.95 & 0.00 & 0.03 & 0.90\\
5000 & 0.00 & 0.19 & 0.94 & 0.00 & 0.07 & 0.93\\
10000 & 0.00 & 0.05 & 0.94 & 0.00 & 0.10 & 0.92\\[2pt]
\hdashline\noalign{\vskip 1ex}
\multicolumn{7}{l}{Partial TMLE} \\[2pt]
500 & 0.01 & 0.18 & 0.93 & 0.00 & 0.01 & 0.92\\
1000 & 0.01 & 0.19 & 0.95 & 0.00 & 0.05 & 0.93\\
5000 & 0.00 & 0.19 & 0.94 & 0.00 & 0.07 & 0.91\\
10000 & 0.00 & 0.11 & 0.94 & 0.00 & 0.08 & 0.93\\
\bottomrule
\end{tabular}
\end{table}

As in the previous Table, we also see in Table \ref{tab:simthetat} that, as expected, the estimators are unbiased. Confidence interval coverage is worse than in Table \ref{tab:simtheta} under the smaller sample sizes of $n=500$ and $n=1000$ for the transported estimators. %, reflecting that more is being asked of the data in this transported scenario. 
However, coverage improves and is close to 95\% in the larger sample sizes of $n=5000$ and $n=10000$. Performance of the one-step and partial TML estimators were similar in these simulated scenarios.

\begin{table}[H]
\centering
\caption{Simulation results comparing one-step and partial TML estimators for the transported interventional indirect and direct effects under two data-generating mechanisms: one with a single, binary $M$ and a single, binary $Z$; and another with multiple, binary $M$s and multiple, binary $Z$s.}
\footnotesize
\label{tab:simthetat}
\begin{tabular}{rcccccc}
\toprule
\multicolumn{1}{c}{ } & \multicolumn{3}{c}{Direct effect} & \multicolumn{3}{c}{Indirect effect} \\
\cmidrule(l{3pt}r{3pt}){2-4} \cmidrule(l{3pt}r{3pt}){5-7}
$n$ & $|\text{Bias}|$ & $\sqrt{n} \times |\text{Bias}|$ & 95\% Cov. & $|\text{Bias}|$ & $\sqrt{n} \times |\text{Bias}|$ & 95\% Cov.\\
\midrule
\multicolumn{7}{l}{\textbf{Binary $M$ and $Z$}} \\[2pt]
\hdashline\noalign{\vskip 1ex}
\multicolumn{7}{l}{One-step} \\[2pt]
500 & 0.00 & 0.03 & 0.86 & 0.00 & 0.01 & 0.89\\
1000 & 0.00 & 0.13 & 0.90 & 0.00 & 0.03 & 0.92\\
5000 & 0.00 & 0.02 & 0.95 & 0.00 & 0.01 & 0.94\\
10000 & 0.00 & 0.09 & 0.95 & 0.00 & 0.04 & 0.94\\[2pt] 
\hdashline\noalign{\vskip 1ex}
\multicolumn{7}{l}{Partial TMLE} \\[2pt]
500 & 0.00 & 0.01 & 0.83 & 0.00 & 0.06 & 0.90\\
1000 & 0.01 & 0.26 & 0.89 & 0.00 & 0.09 & 0.93\\
5000 & 0.00 & 0.09 & 0.96 & 0.00 & 0.02 & 0.95\\
10000 & 0.00 & 0.05 & 0.95 & 0.00 & 0.01 & 0.95\\
\midrule
\multicolumn{7}{l}{\textbf{Multivariate $M$ and $Z$}} \\[2pt]
\hdashline\noalign{\vskip 1ex}
\multicolumn{7}{l}{One-step} \\[2pt]
500 & 0.01 & 0.11 & 0.93 & 0.00 & 0.01 & 0.91\\
1000 & 0.00 & 0.09 & 0.95 & 0.00 & 0.03 & 0.93\\
5000 & 0.00 & 0.18 & 0.95 & 0.00 & 0.07 & 0.91\\
10000 & 0.00 & 0.08 & 0.95 & 0.00 & 0.03 & 0.95\\[2pt]
\hdashline\noalign{\vskip 1ex}
\multicolumn{7}{l}{Partial TMLE} \\[2pt]
500 & 0.00 & 0.08 & 0.93 & 0.00 & 0.01 & 0.92\\
1000 & 0.01 & 0.25 & 0.93 & 0.00 & 0.01 & 0.89\\
5000 & 0.00 & 0.12 & 0.95 & 0.00 & 0.07 & 0.93\\
10000 & 0.00 & 0.09 & 0.94 & 0.00 & 0.02 & 0.94\\
\bottomrule
\end{tabular}
\end{table}

% \begin{table}[H]
% \centering
% \caption{Multivariate M and Z.}
% \footnotesize
% \begin{tabular}{rcccccc}
% \toprule
% \multicolumn{1}{c}{ } & \multicolumn{3}{c}{Direct effect} & \multicolumn{3}{c}{Indirect effect} \\
% \cmidrule(l{3pt}r{3pt}){2-4} \cmidrule(l{3pt}r{3pt}){5-7}
% $n$ & $\text{Bias}$ & $\sqrt{n} \times \text{Bias}$ & 95\% Cov. & $\text{Bias}$ & $\sqrt{n} \times \text{Bias}$ & 95\% Cov.\\
% \midrule
% \multicolumn{7}{l}{\textit{Not transported}} \\[2pt]
% 500 & 0.01 & 0.18 & 0.95  & 0.00 & 0.01 & 0.89 \\
% 1000 & 0.00 & 0.14 & 0.92 & 0.00 & 0.03 & 0.88 \\
% 5000 & 0.00 & 0.15 & 0.94 & 0.00 & 0.01 & 0.94 \\
% 10000 & 0.00 & 0.07 & 0.97 & 0.00 & 0.02 & 0.94 \\
% \midrule
% \multicolumn{7}{l}{\textit{Transported}} \\[2pt]
% 500 & 0.02 & 0.35 & 0.90   & 0.00 & 0.01 & 0.85 \\
% 1000 & 0.02 & 0.46 & 0.92  & 0.00 & 0.02 & 0.88 \\
% 5000 & 0.00 & 0.04 & 0.94  & 0.00 & 0.04 & 0.92 \\
% 10000 & 0.00 & 0.09 & 0.93 & 0.00 & 0.09 & 0.93 \\
% \bottomrule
% \end{tabular}
% \end{table}

% Table \ref{} shows simulation results
% \begin{table}[H]
% \footnotesize
% \centering
% \begin{tabular}[t]{llccc}
% \toprule
% $n$ & Estimand & $|\text{Bias}|$ & $\sqrt{n} \times |\text{Bias}|$ & 95\% CI cov.\\
% \midrule
% 500 & Direct & 0.02 & 0.44 & 0.93\\
% 500 & Indirect & 0.02 & 0.37 & 0.86\\
% 1000 & Direct & 0.01 & 0.30 & 0.91\\
% 1000 & Indirect & 0.01 & 0.37 & 0.89\\
% 5000 & Direct & 0.00 & 0.08 & 0.94\\
% 5000 & Indirect & 0.00 & 0.23 & 0.91\\
% 10000 & Direct & 0.00 & 0.00 & 0.96\\
% 10000 & Indirect & 0.00 & 0.17 & 0.94\\
% \bottomrule
% \end{tabular}
% \end{table}

\section{Illustrative Example}
\subsection{Background}
We now apply these estimators to a reanalysis of the Moving to Opportunity (MTO) study, which was briefly described in the Introduction. In MTO, families were randomized to receive a Section 8 (also called Housing Choice) voucher that they could use to move out of public housing and into a rental on the private market \citep{sanbonmatsu2011moving}. Previously, others found evidence suggesting differences across MTO sites in the effect of housing voucher receipt on later behavioral problems (using the Behavioral Problems Index (BPI) \citep{zill1990behavior}) among girls in adolescence \citep{osypuk2012differential}. It may be of interest to examine whether these site differences are also seen for mediational indirect effects, and if so, if there are certain mediators in particular that likely drive such differences. %In this study, boys in families that were randomized to receive the voucher developed psychiatric disorders (diagnosed following the Diagnostic and Statistical Manual, Version IV (DSM-IV)) at higher rates than boys in families that were randomized to not receive the voucher \citep{kessler2014associations, sanbonmatsu2011moving,kling2007experimental,rudolph2021helped}. 
%We have been interested in understanding why these and other, related, unintended negative effects occurred in MTO. 
Previously, we hypothesized relevant mediators linking voucher receipt and use with adolescent mental health and risk behavior outcomes include: 1) aspects of the school environment (i.e., school ranking, Title I status, etc.), 2) number of residential and school moves, and 3) neighborhood poverty, and we previously considered such mediatiors together as a bundle \citep{rudolph2021helped}. 

\begin{figure}[!ht]
       \caption{Directed acyclic graphs (DAGs) considering each mediator grouping separately in the Moving to Opportunity Study. Bolded text indicates mediator groupings. Red arrows indicate the indirect effect path.}
     \label{fig:dag}
     \centering
    \subfloat[DAG considering features of the school environment as the mediators. \label{fig:dag:a}]{
    \includegraphics[width=.45\textwidth]{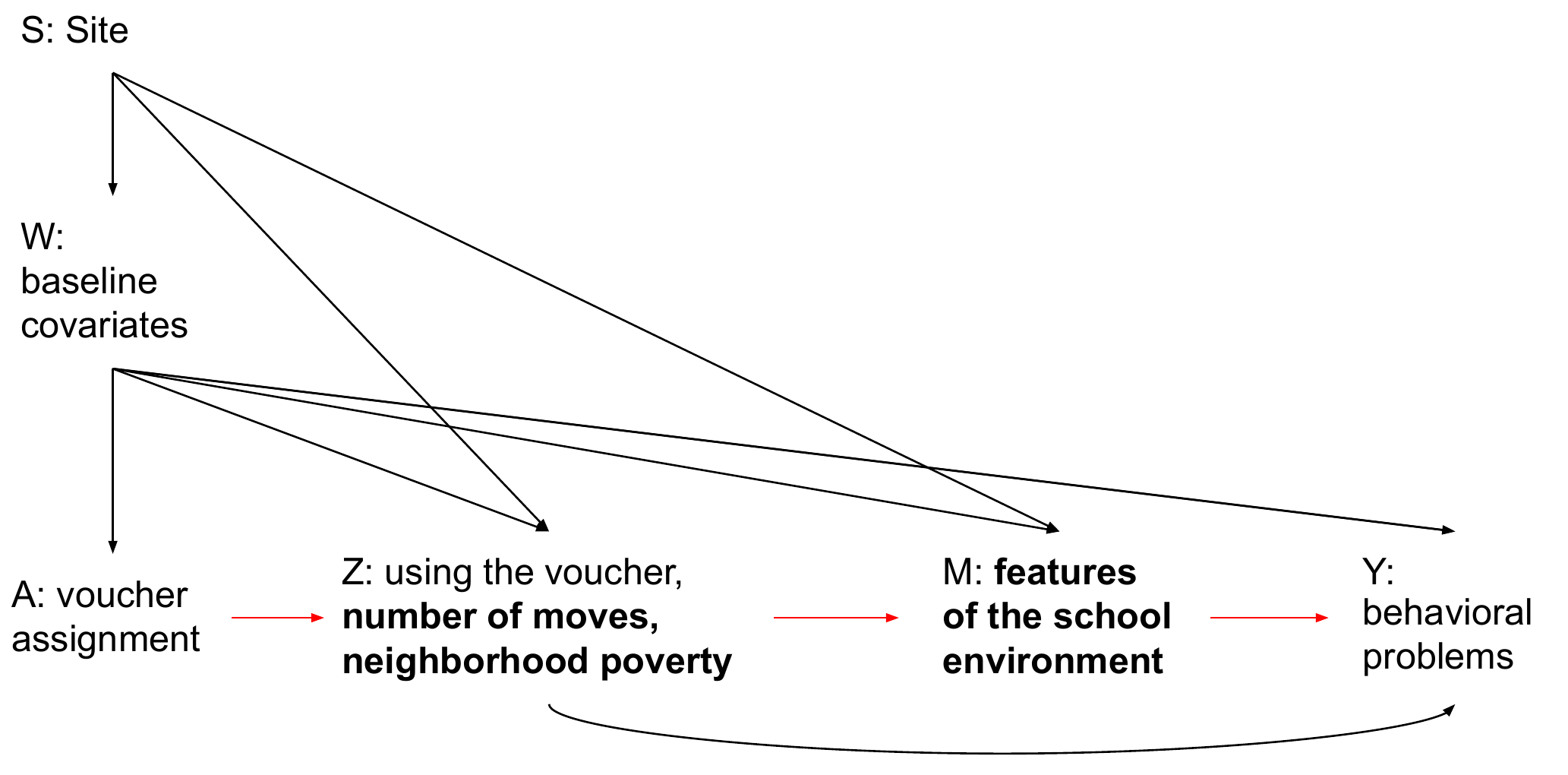}}
    \hfill
    \centering
    \subfloat[DAG considering neighborhood poverty as the mediator. \label{fig:dag:b}]{
    \includegraphics[width=.45\textwidth]{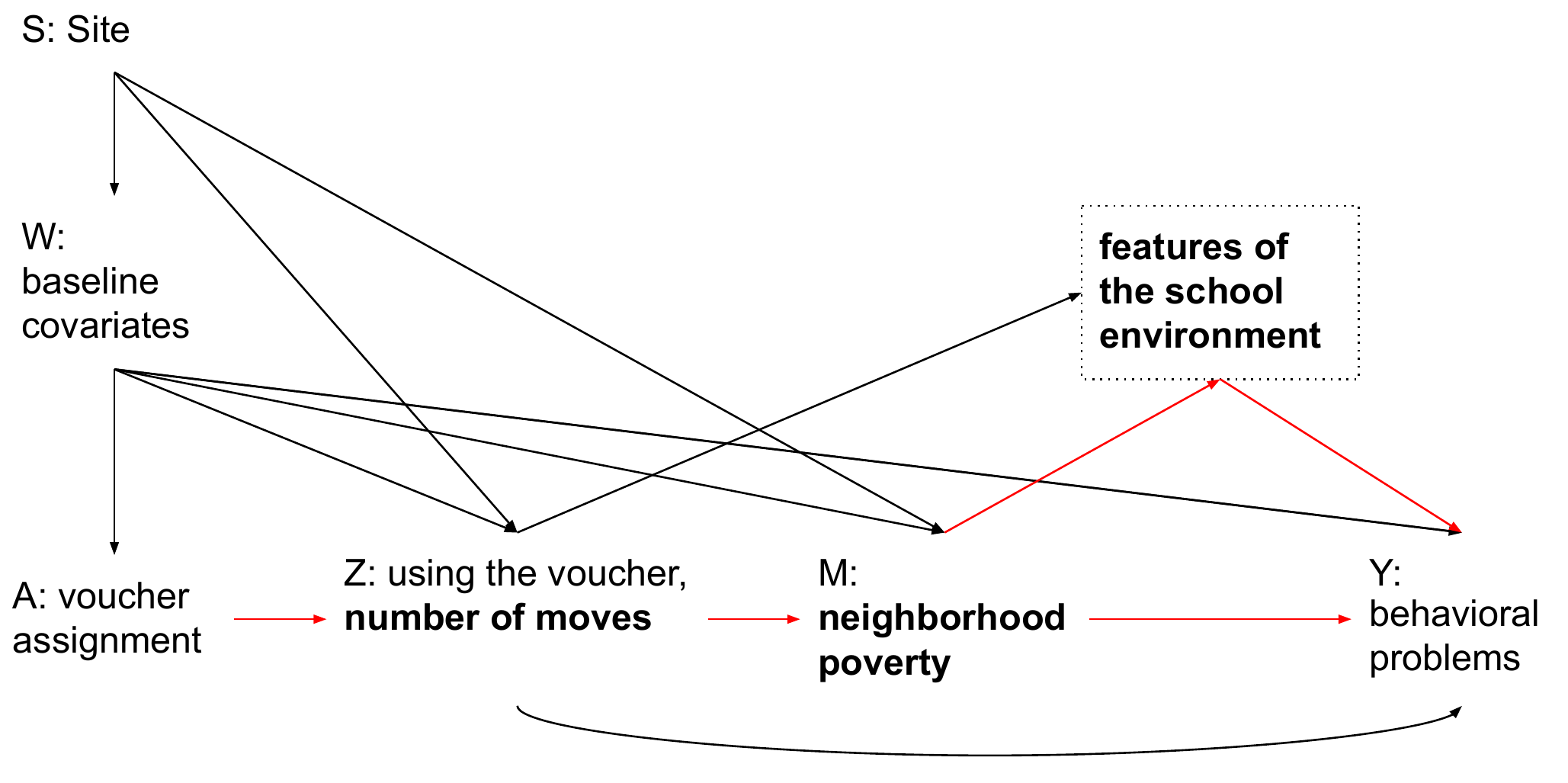}}
    \hfill
\\
\centering \subfloat[DAG considering number of residential and school moves as the mediators. \label{fig:dag:c}]{
    \includegraphics[width=.45\textwidth]{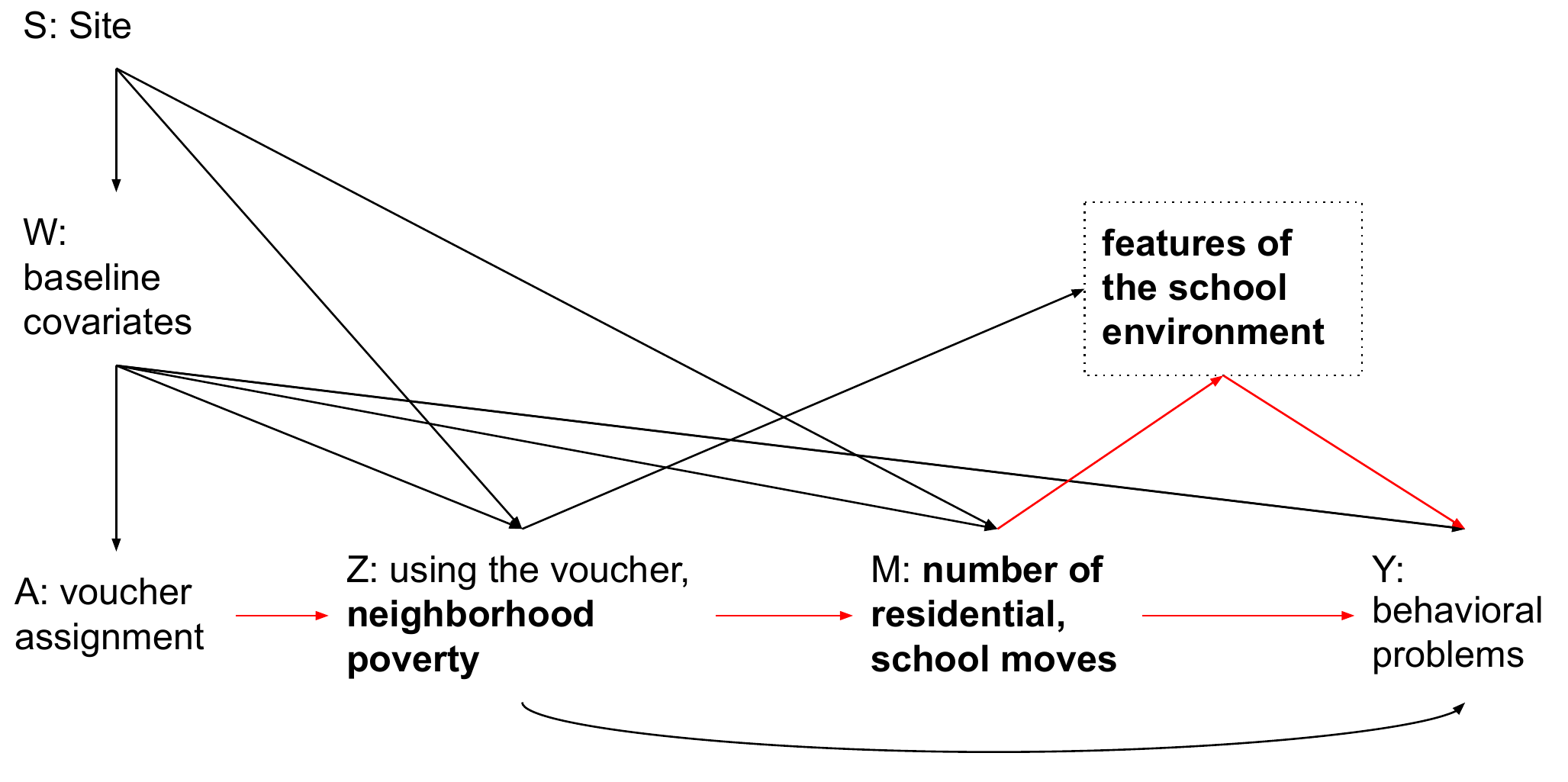}}
   \end{figure}

However, it may also be of interest to estimate the indirect effect that operates through each of the above mediator groupings separately. When we do that---for example, when we estimate the indirect effect through the school environment---we need to consider whether the other mediators, the subset that we are not examining, should be treated as post-treatment intermediate confounders or not. %whether they should be relegated to the direct effect path. 
  In many cases, it would be appropriate to treat the subset of mediators we are not examining as post-treatment intermediate confounders. For example, consider the indirect effect of voucher receipt on behavioral problems %development of a psychiatric disorder 
in adolescence that operates through features of the school environment. The subset of original mediators that we are no longer examining includes number of residential and school moves and neighborhood poverty. Each of these is affected by voucher receipt and may affect aspects of the school environment and risk of developing a psychiatric disorder, so meets the definition of an intermediate confounder (see Figure \ref{fig:dag:a}). Consequently, we would want to control for them as part of $Z$, as they represent potential confounders of the $M-Y$ relationship. 

However, there are other cases where it would not be appropriate to treat the subset of original mediators that we are no longer examining as intermediate confounding variables. As another example, consider the indirect effect of voucher receipt on behavioral problems %development of a psychiatric disorder 
in adolescence that operates through neighborhood poverty. In that case, the subset of mediators we are no longer examining includes the number of residential and school moves and features of the school environment. The number of residential and school moves is affected by voucher receipt and may affect neighborhood poverty, so meets the definition of an intermediate confounder. However, features of the school environment, while affected by voucher receipt, would be \textit{affected by} neighborhood poverty. Thus, features of the school environment would not meet the definition of an intermediate confounder (see Figure \ref{fig:dag:b}), so we should not include them as part of $Z$. Conditioning on features of the school environment may bias estimation of the $M-Y$ relationship, because they lie on the causal pathway from $M$ to $Y$. In addition, because features of the school environment are affected by both $Z$ and $M$, conditioning on them would open a backdoor pathway between $M$ and $Y$ and between $Z$ and $Y$. Consequently, we would not adjust for features of the school environment in this analysis, which would result in these variables contributing to %By ignoring features of the school environment in the analysis, it would be contribute 
to two different paths: 1) the path $A \rightarrow Z \rightarrow \text{school environment} \rightarrow Y$, subsumed into the direct effect, and 2) the path $A \rightarrow Z \rightarrow M  \rightarrow \text{school environment} \rightarrow Y,$ subsumed into the indirect effect. 

\subsection{Statistical Analysis}
We used our proposed nontransport estimators to estimate the IIE and IDE through each of the above mediator groupings separately: 1) through features of the school environment, treating neighborhood poverty and number of residential and school moves as intermediate confounders; 2) through neighborhood poverty, treating number of residential and school moves as intermediate confounders; and 3) through number of residential and school moves, treating neighborhood poverty as the intermediate confounder. 

As stated above, we included only adolescent girls in participating MTO families. We excluded the Baltimore site, because a different housing voucher program was implemented in this city concurrently. This resulted in a rounded sample size of N=2,200. We list and define each variable used in the analysis in Section A2 of the online appendix. %, including the amount of missingness.
For simplicity, and because this analysis was for illustrative purposes, we used one imputed dataset, imputed using multiple imputation by chained equations \citep{buuren2010mice}.

We then use our proposed transport estimators to estimate transported direct and indirect effects, through each of the above mediator groupings separately, transporting from Los Angeles to Boston. Again, we included adolescent girls in families who participated in MTO in the Los Angeles (rounded sample size n=550) and Boston (rounded sample size n=400) sites. We used the same variables and imputation as described above.

We implemented both the nontransport and transport estimators using 10 folds for cross-fitting. We used Super Learner \citep{van2007super} to fit the nuisance parameters, implemented using the SuperLearner package \citep{polley2017super}, including the following algorithms: intercept-only regression, generalized linear regression, lasso \citep{tibshirani1996regression}, multiple additive regression splines \citep{friedman1991multiple}, and gradient boosted machines \citep{chenXGBoost}. Columbia University determined this analysis of deidentified data to be non-human subjects research. %Code to replicate our analyses is available: \url{blinded for review}.

\subsection{Results}
\begin{figure}[H]
       \caption{Inverventional indirect and direct effect point estimates and 95\% confidence intervals of the effect of Section 8 housing voucher receipt on later behavioral problems index score in adolescence among girls in the Moving to Opportunity study. Indirect effects are those operating through mediators (all mediators considered together as a bundle and separately by mediator group) and direct effects are those not operating through mediators. Estimates are given separately for the one-step and partial TML estimators. All results were approved for release by the U.S. Census Bureau, authorization number 
CBDRB-FY23-CES018-005. }
     \label{fig:nt}
     \centering
    \subfloat[Interventional indirect effects. \label{fig:ntindirect}]{
    \includegraphics[width=.45\textwidth]{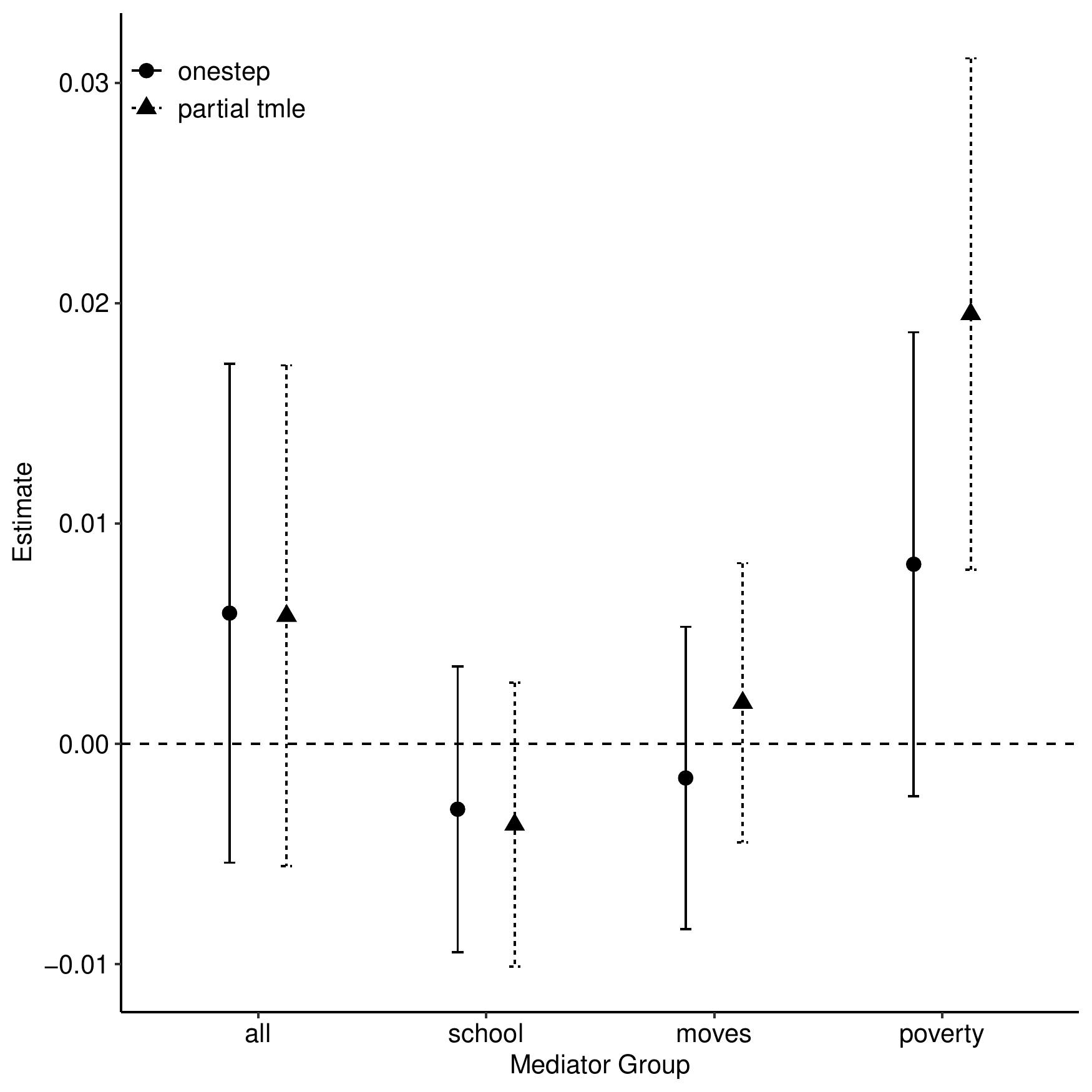}}
    \hfill
    \centering
    \subfloat[Interventional direct effects.  \label{fig:ntdirect}]{
    \includegraphics[width=.45\textwidth]{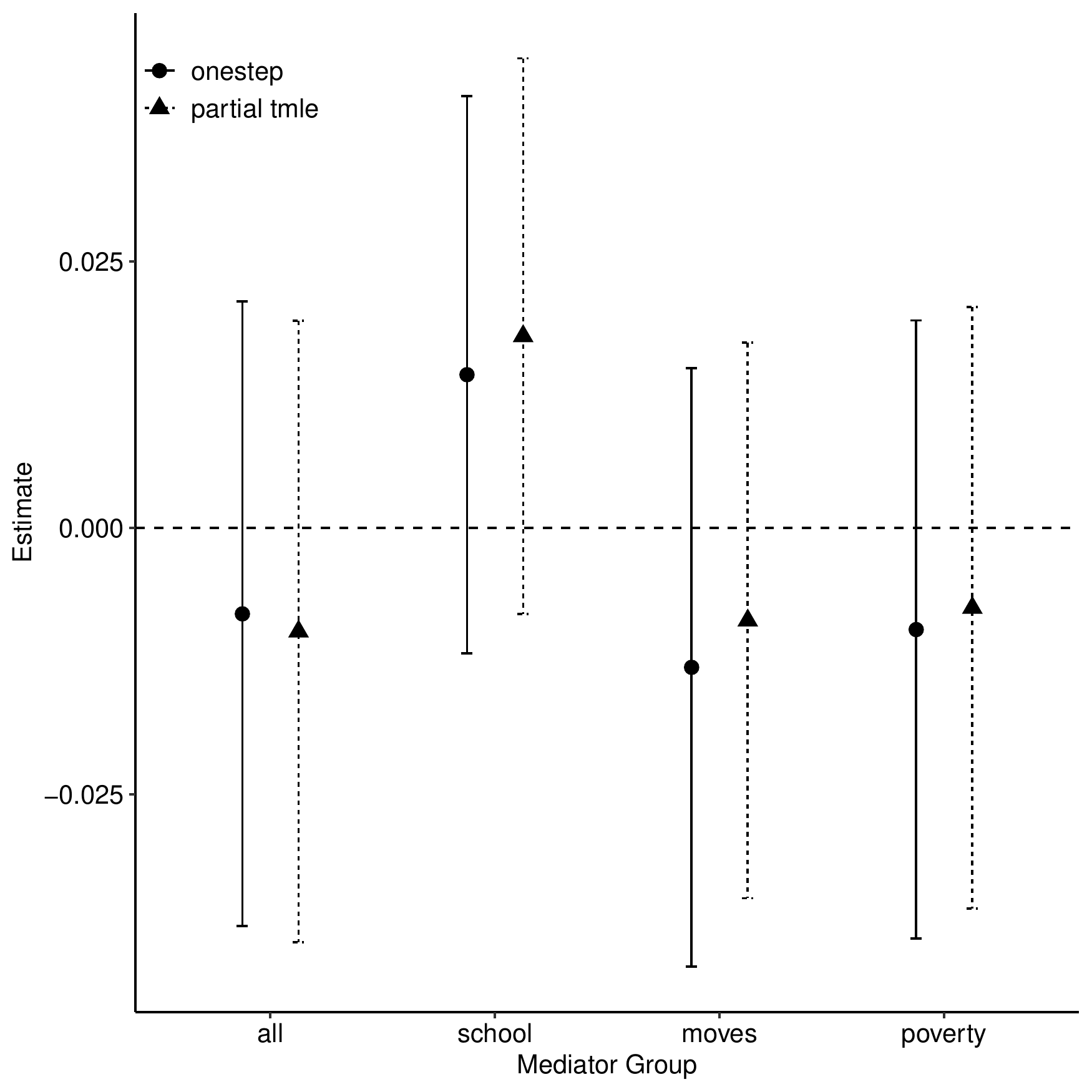}}
   \end{figure}

First, Figure \ref{fig:nt} shows the point estimates and 95\% CIs for the nontransported IIE (Figure \ref{fig:ntindirect}) and IDE (Figure \ref{fig:ntdirect}) across all sites (Boston, Chicago, Los Angeles, and New York) for all mediators considered together as a bundle as well as separately by each mediator grouping. 

Averaging all sites together, we see that there is a slight harmful, though nonsignificant, indirect effect of voucher receipt on behavioral problems in adolescence through mediators related to the school and neighborhood environments as well as residential and school instability (difference in BPI score (ATE): 0.006, 95\% CI: -0.005, 0.017 for the one-step estimator). Separating out the indirect effects by mediator grouping, we see that the indirect effect through neighborhood poverty is harmful, and significantly so in the case of the partial TML estimator (ATE: 0.020, 95\% CI: 0.008, 0.031). In contrast, it appears that the indirect effect through features of the school environment is slightly beneficial and may contribute to a reduction in behavioral problems, though these estimates are not significant (ATE: -0.003, 95\% CI: -0.009, 0.004) for the one-step estimator. The indirect effects through number of residential and school moves appears null.

\begin{figure}[H]
       \caption{Transported and non-transported inverventional indirect and direct effect point estimates and 95\% confidence intervals of the effect of Section 8 housing voucher receipt on later behavioral problems index score in adolescence among girls in the Moving to Opportunity study. Indirect effects are those operating through mediators (all mediators considered together as a bundle and separately by mediator group) and direct effects are those not operating through mediators. Estimates are given only for the  one-step estimator in parts A and B, as partial TMLE estimates were similar. All results were approved for release by the U.S. Census Bureau, authorization number 
CBDRB-FY23-CES018-005. }
     \label{fig:t}
     \centering
    \subfloat[Interventional indirect effects. \label{fig:tindirect}]{
    \includegraphics[width=.45\textwidth]{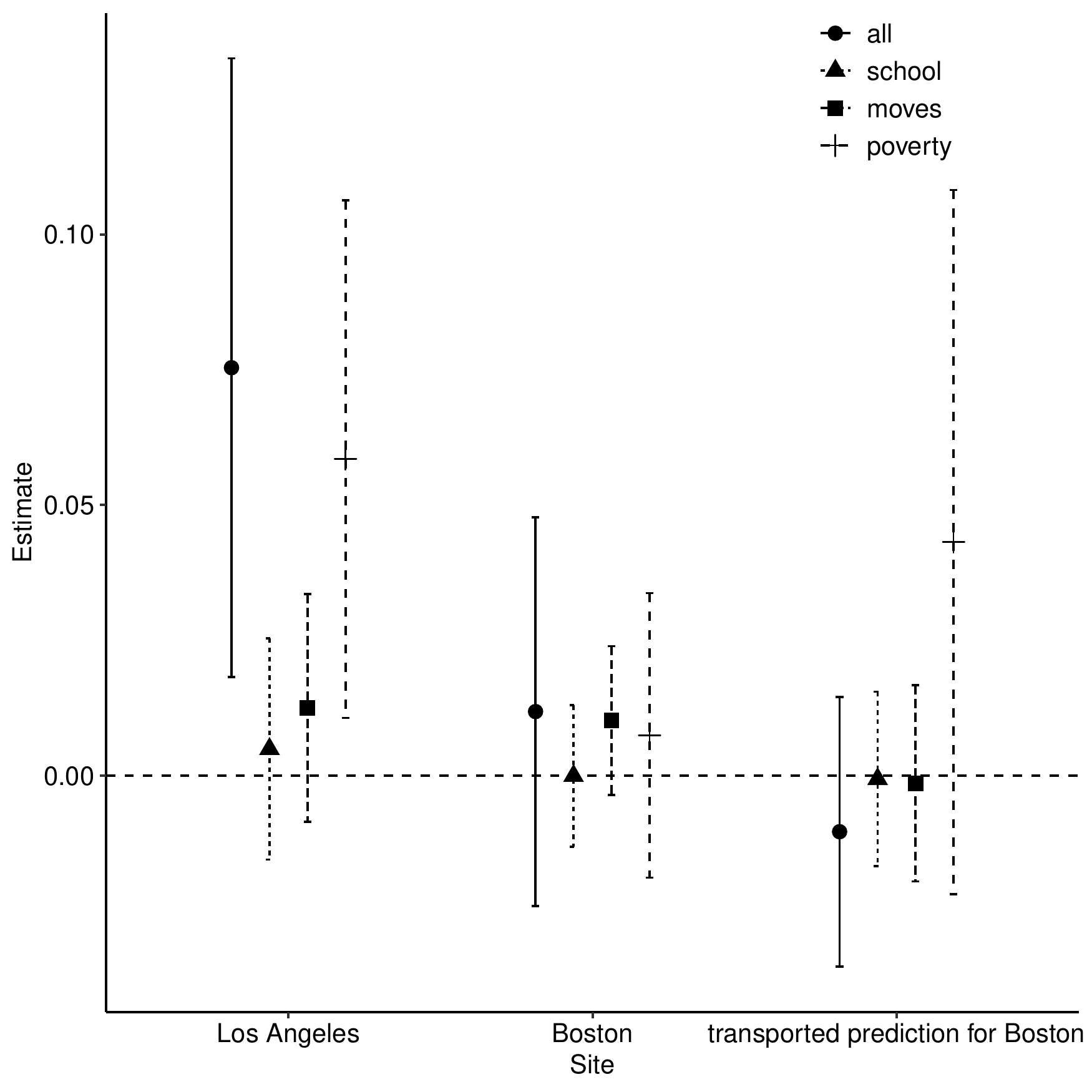}}
    \hfill
    \centering
    \subfloat[Interventional direct effects.  \label{fig:tdirect}]{
    \includegraphics[width=.45\textwidth]{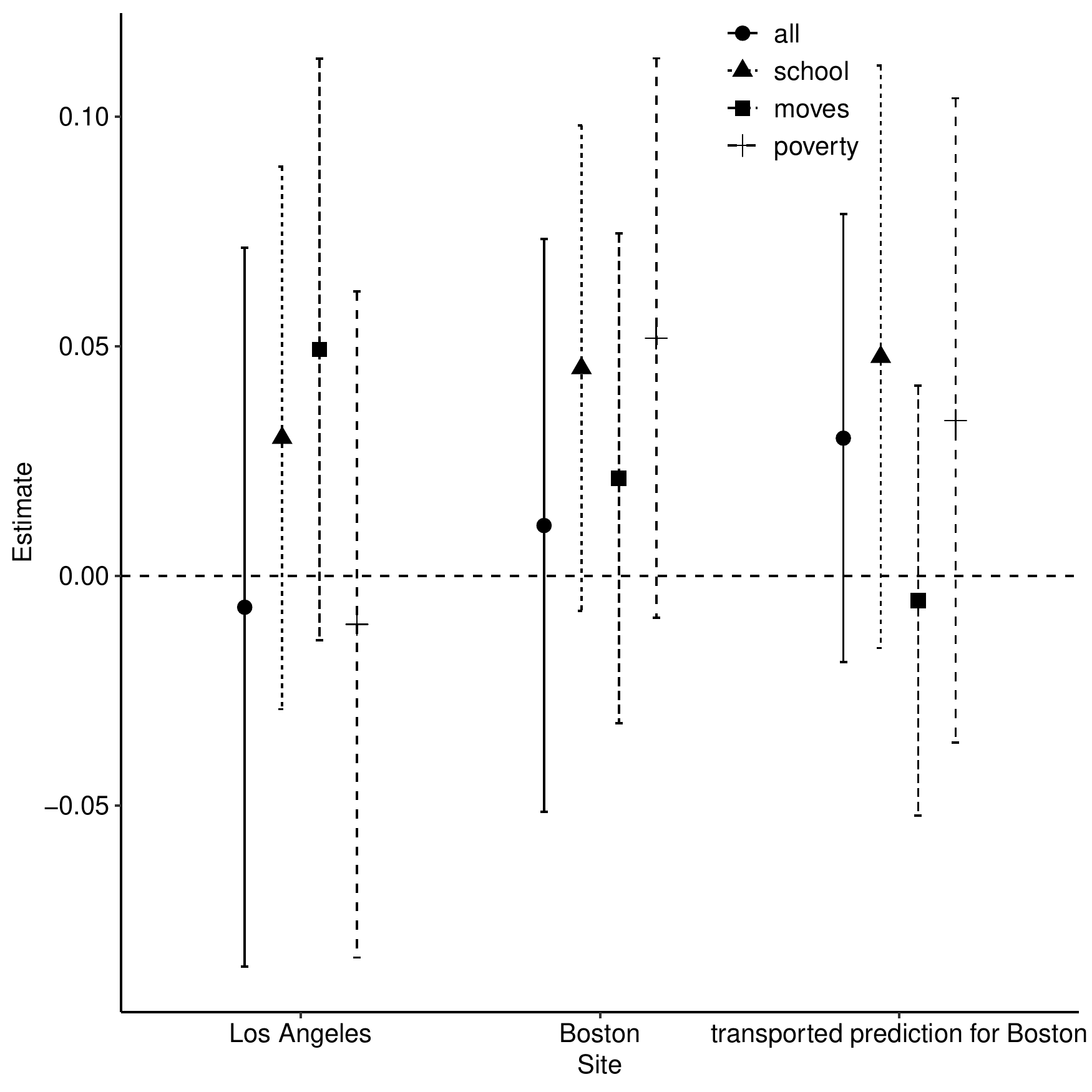}}
    \\
    \centering
    \subfloat[Transported and non-transported interventional indirect effects through neighborhood poverty comparing estimates from the one-step estimator to those from partial TMLE.  \label{fig:tindirectpoverty}]{
\includegraphics[width=.45\textwidth]{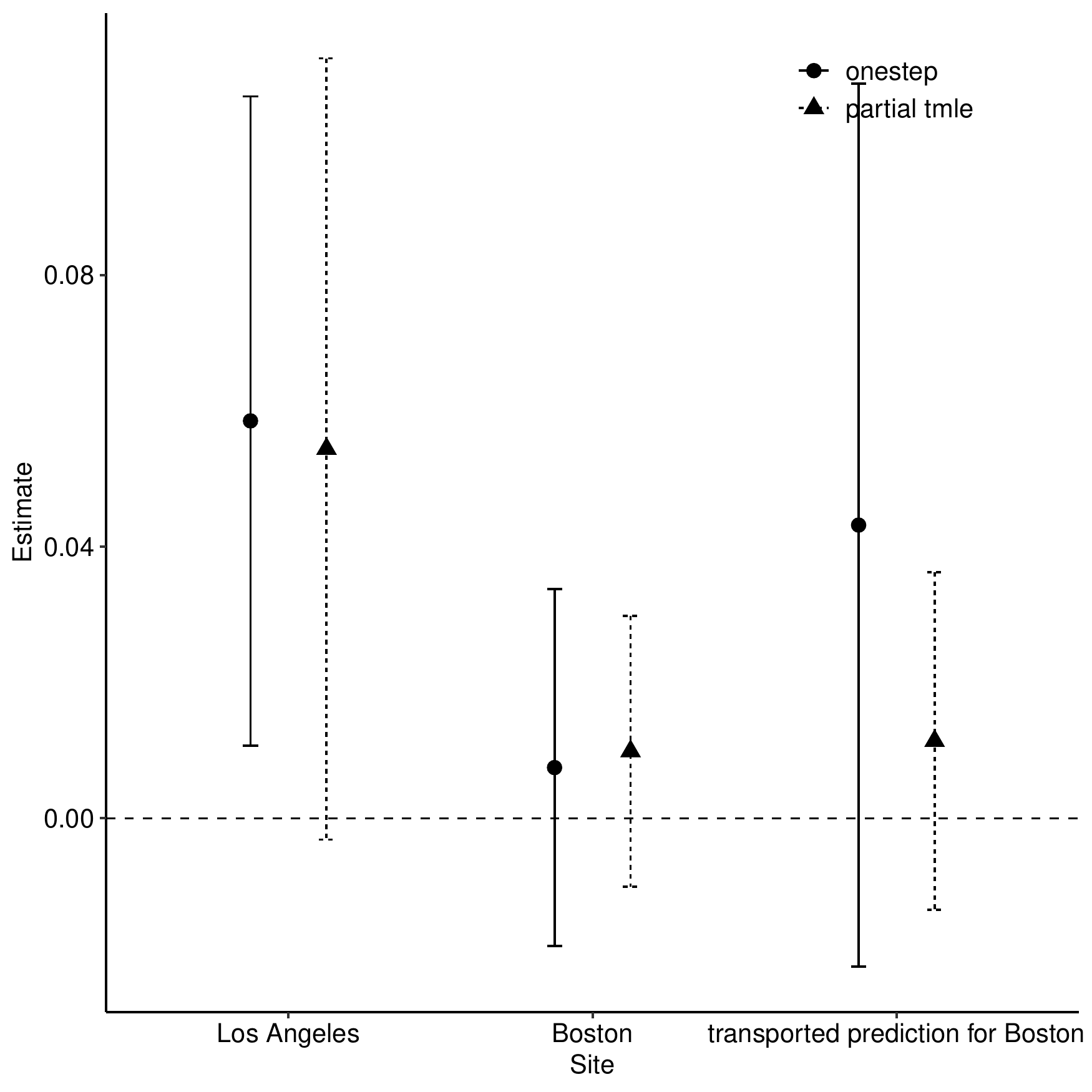}}

  \hfill
   \end{figure}

We now turn our attention to site-specific estimates of these effects, including transported estimates. We see in Figure \ref{fig:tindirect} a substantial difference in the point estimates of the indirect effect of housing voucher receipt on behavioral problems through all considered mediators, comparing Los Angeles to Boston, in which the behavioral problems of adolescent girls in Los Angeles seem negatively affected by the intervention whereas the behavioral problems of adolescent girls in Boston do not seem affected. We note that we only shown estimates from the one-step estimator here, as estimates from the partial TMLE were generally similar, with an exception discussed further below. Our transport estimator accounts for differences in the distribution of baseline covariates, $W$, differences in the conditional distribution of intermediate confounding variables, $Z$, and differences in the conditional distribution of mediating variables, $M$. We see that accounting for these differences largely explains the initial difference in point estimates when all mediators are considered together as a bundle as well as when mediators related to the school environment are considered. When considering mediators of the number of residential moves and number of school moves, there is no initial difference in indirect effects between the two sites. 

When considering neighborhood poverty as the sole mediator of the indirect effect, the transported point estimate predicted for Boston is closer to the observed point estimate for Los Angeles than it is to the observed point estimate for Boston and has wide confidence intervals. This is one case where the one-step and partial TMLE estimates differed, as shown in Figure \ref{fig:tindirectpoverty}. Using the partial TML estimator that accounts for differences in the distribution of $W, Z, $ and $M$, the transported indirect effect estimate for Boston is close to Boston's observed indirect effect estimate. When considering neighborhood poverty as the sole mediator, practical violations of the positivity assumption were more of an issue: 15\% of observations had a predicted ratio $\frac{1-\s(1,z,m,w)}{\s(1,z,m,w)} > 100$ and a 75th percentile value of 42, as compared to 12\% of observations with a predicted ratio $\frac{1-\s(1,z,m,w)}{\s(1,z,m,w)} > 100$ and a 75th percentile value of 30 when all mediators were considered together as a bundle. We would expect the partial TMLE to return more stable estimates in such a setting.

\section{Discussion}
In this paper, we proposed doubly robust, data-adaptive estimators---a one-step estimator and a partial TML estimator---of interventional direct and indirect effects and transported interventional direct and indirect effects that accommodate multivariate mediators and multivariate intermediate confounding variables simultaneously. This represents a step forward in allowing these effects to be estimated in real-world data, because multiple mediators and multiple intermediate confounders frequently exist in practice. In addition to proposing novel estimators, we also walk through a strategy for separating indirect effects into mediator- or mediator-group-specific indirect effects, while appropriately accounting for other, possibly co-occurring intermediate variables. 

Although we addressed a limitation of much previous mediation work here, other limitations remain. First, our proposed estimators involve estimation of numerous nuisance parameters. Although not all parameters need to be estimated consistently for the estimator to be consistent due to double robustness, they do all need to be estimated consistently for the resulting variance estimates to be accurate \citep{benkeser2017doubly}. 

Second, certain pairs of the nuisance parameters should be congenial with each other \citep{dukes2019doubly,vansteelandt2011instrumental}, e.g., $\g$ and $\e$, $\q$ and $\rr$. Although consistency---not congeniality---is required for estimator consistency, the two are related in that consistency may be harder to achieve without congeniality. It is plausible that lack of congeniality may affect the estimator's finite sample performance \citep{dukes2019doubly,vansteelandt2011instrumental}. However, there is no clear way to achieve such congeniality with the data-adaptive model-fitting approach we propose. Understanding practical consequences of non-congenial nuisance parameter estimation and approaches to achieve congeniality in data-adaptive estimation are areas ripe for future work. 

Third, positivity proved problematic in the illustrative example, particularly for $\theta^T,$ due to small predicted probabilities in $\hat \s(a,z,m,w)$. This problem may be considered a ``catch-22'', meaning a situation that involves mutually dependent conditions that are in conflict.  In order for the transport assumption to hold, we need to include all variables required for $S$-admissibility \citep{pearl2015generalizing}. If those variables do not just include baseline covariates, $W$, then $\E(Y \mid A, W, S) \ne \E(Y \mid A, W). $ If additional variables are added to achieve $S$-admissibility such that $\E(Y \mid M, Z, A, W, S) = \E(Y \mid M, Z, A, W),$ then that means that those additional variables explain the ways in which the outcome model differs across sites. But, the explanatory power of these variables comes at the price of practical violations of the positivity assumption; meaning, that certain values of these additional variables, $m, z$, in combination with the baseline covariates and treatment, make the selection mechanism for site close to deterministic. This points to another area for future work on transport estimation: to pursue estimation strategies that are more robust to practical positivity violations, such as the partial TMLE we proposed here, and to pursue alternative estimands that rely on a weaker positivity assumption.%The componenets of the positivity assumption that are particularly problematic are: $\p(w \mid S=0)>0, \p(z \mid a',w, S=0)>0$, and $\p(m \mid a^\star, w, S=0)>0$ imply $\p(z \mid a',w, S=1)>0$ and $\p(m \mid z,a',w \mid S=1)>0 $.

Lastly, our proposed estimators, particularly, the transport estimators, may not perform well in small, finite samples. This is evidenced, in part, by undercoverage even in the relatively simple simulation settings considered for the transported IIE and IDE in sample sizes of $n=500$ and $n=1000$. Consequently, we are exploring strategies to improve efficiency at the expense of making a one or a limited number of reasonable assumptions on the statistical model.

\vspace{2cm}
\noindent \textbf{Acknowledgements:} This research was conducted as a part of the U.S. Census Bureau's Evidence Building Project Series. The Census Bureau has reviewed this data product to ensure appropriate access, use, and disclosure avoidance protection of the confidential source data used to produce this product (Data Management System (DMS) number:  P-7504667, Disclosure Review Board (DRB) approval number:  CBDRB-FY23-CES018-005). 

This work was supported by R01DA053243 (PI Rudolph).

\begin{appendix}
\renewcommand{\thesection}{A\arabic{section}}

\section{Simulation data generating mechanisms}

\subsection{Binary $M$ and $Z$: $\theta$}

Indirect effect = 0.0975; direct effect = 0.1933; indirect effect efficiency bound = 0.3191; direct effect efficiency bound = 1.4607.

{
\footnotesize
\begin{align*}
\P(W=1) &= 0.4 \\
\P(A=1) &= 0.5 \\
\P(Z=1 \mid A,W) &= \expit(-\log(2) + \log(10)A - \log(2)W) \\
\P(M=1 \mid A,Z,W) &= \expit(-\log(2) + \log(12)Z - \log(1.4)W \\
\P(Y=1 \mid A,Z,M,W) &= \expit(-\log(5) + \log(8)Z + \log(10)M - \log(1.2)W + \log(1.2)ZW)
\end{align*}
}

\subsection{Binary $M$ and $Z$: $\theta^T$}

Indirect effect = 0.0522; direct effect = 0.1347; indirect effect efficiency bound = 0.019; direct effect efficiency bound = 0.1742

{
\footnotesize
\begin{align*}
\P(W=1) &= 0.4 \\
\P(A=1) &= 0.5 \\
\P(Z=1 \mid A,W,S) &= \expit(-\log(2) + \log(4)A - \log(2)W) + \log(1.4)S \\
\P(M=1 \mid A,Z,W,S) &= \expit(-\log(2) + \log(10)Z - \log(1.4)W + \log(0.3)S\\
\P(Y=1 \mid A,Z,M,W) &= \expit(-\log(5) + \log(8)Z + \log(6)M - \log(1.2)W + \log(1.2)ZW)
\end{align*}
}

\subsection{Multivariate $M$ and $Z$: $\theta$}

Indirect effect = 0.0177; direct effect = 0.0314; indirect effect efficiency bound = 0.0749; direct effect efficiency bound = 0.9951.

{
\footnotesize
\begin{align*}
\P(W=1) &= 0.4 \\
\P(A=1) &= 0.5 \\
\P(Z1=1 \mid A,W) &= 0.25 + 0.1A + 0.2W \\
\P(Z2=1 \mid A,W) &= 0.4 + 0.1A - 0.1W \\
\P(M1=1 \mid A,Z1,Z1,W) &= 0.6 + 0.1Z1 + 0.05A - 0.3W \\
\P(M2=1 \mid A,Z2,Z2,W) &= 0.33 + 0.22Z2 + 0.05A + 0.15W \\
\P(Y=1 \mid A,Z1,Z1,M1,M1,W) &= \expit(-\log(5) + \log(8)Z1 + \log(4)M1 - \log(1.2)W  - \log(2)Z2 + \\ \log(1.2)M2 + \log(1.2)WZ1)
\end{align*}
}

\subsection{Multivariate $M$ and $Z$: $\theta^T$}

Indirect effect = 0.0177; direct effect = 0.0313; indirect effect efficiency bound = 0.0083; direct effect efficiency bound = 0.1093. 

{
\footnotesize
\begin{align*}
\P(W=1) &= 0.4 \\
\P(A=1) &= 0.5 \\
\P(Z1=1 \mid A,W,S) &= 0.25 + 0.1A + 0.2W + 0.05S \\
\P(Z2=1 \mid A,W,S) &= 0.4 + 0.1A - 0.1W + 0.075S \\
\P(M1=1 \mid A,Z1,Z1,W,S) &= 0.6 + 0.1Z1 + 0.05A - 0.3W \\
\P(M2=1 \mid A,Z2,Z2,W,S) &= 0.33 + 0.22Z2 + 0.05A + 0.15W - 0.05S \\
\P(Y=1 \mid A,Z1,Z1,M1,M1,W) &= \expit(-\log(5) + \log(8)Z1 + \log(4)M1 - \log(1.2)W  - \log(2)Z1 + \\ \log(1.2)M2 + \log(1.2)WZ1)
\end{align*}
}

\section{Illustrative example variable names and definitions}
Baseline covariates included the following (with percent missing given in parentheses):
 \begin{itemize}
     \item Adolescent characteristics (all had 0\% missing except race/ethnicity, which had 2\% missing): site (Boston, Chicago, LA, NYC), age, race/ethnicity (categorized as black, latino/Hispanic, white, other), number of family members 
(categorized as 2, 3, or 4+), someone from school asked to discuss problems the child had with schoolwork or behavior during the 2 years prior to baseline, child enrolled in special class for gifted and talented students.
     \item Adult household head characteristics (which all had 0\% missing): high school graduate, marital status (never vs ever married), whether had been a teen parent, work status, receipt of AFDC/TANF, whether any family member has a disability.
     \item Neighborhood characteristics (which all had 0\% missing except neighborhood poverty, which had 2\% missing): felt neighborhood streets were unsafe at night; very dissatisfied with neighborhood; poverty level of neighborhood.
     \item Reported reasons for participating in MTO (which had 0\% missing): to have access to better schools.
     \item Moving-related characteristics (which had 0\% missing): moved more then 3 times during the 5 years prior to baseline, previous application for Section 8 voucher.
 \end{itemize}

The treatment was a binary indicator of whether or not the family was randomized to receive a Section 8 housing voucher (0/1) and had zero missingness. 

Whether or not the family used the voucher to move (0/1) within the 90 days allotted, which had zero missingness.

Mediators/ intermediate confounders, depending on the analysis, included the following, all were duration weighted over the 10-15 year follow up: 
\begin{itemize}
    \item Neighborhood poverty (0\% missing);
    \item Number of residential moves (0\% missing); number of schools (8\% missing);
    \item Characteristics of the school environment:
    \begin{itemize}
        \item School rank (12\% missing),
        \item Title I status (8\% missing),
    \end{itemize}
\end{itemize}

The outcome was score on the Behavioral Problems Index \citep{zill1990behavior}, considered as a continuous variable (13\% missing).

\end{appendix}
\bibliographystyle{biom}
\bibliography{refs}
\end{document}